\documentclass[a4,useAMS,usenatbib,usegraphicx]{mn2e}
\def\apss{\ref@jnl{Ap\&SS}} 
\def\aapr{\ref@jnl{A\&A~Rev.}}

\usepackage{amsmath}
\include{epsf}
\usepackage{float}
\usepackage{bbding}

\usepackage{epsf}
\usepackage{appendix}
\epsfclipon
\title[Ursa Major II Disruption]{Ursa Major II - Reproducing the observed properties through tidal disruption}
\author[R.Smith et al]{R.Smith$^{1}$\thanks{E-mail:rsmith@astro-udec.cl}, M. Fellhauer${^1}$, G. N. Candlish${^1}$, R. Wojtak$^2$, J. P. Farias${^1}$, M. Bla\~{n}a${^1}$ \\
$^{1}$Departamento de Astronomia, Universidad de Concepcion, Casilla 160-C, Concepcion, Chile\\
$^{2}$Dark Cosmology Centre, Niels Bohr Institute, University of Copenhagen, Juliane Maries Vej 30, DK-2100 Copenhagen \O, Denmark}
\begin{document}

\date{Accepted to MNRAS 23rd May 2013}

\pagerange{\pageref{firstpage}--\pageref{lastpage}} \pubyear{2011}

\maketitle

\label{firstpage}

\begin{abstract}
Recent deep photometry of the dwarf spheroidal Ursa Major II's morphology, and spectroscopy of individual stars, have provided a number of new constraints on its properties. With a velocity dispersion $\sim$6~km~s$^{-1}$, and under the assumption that the galaxy is virialised, the mass-to-light ratio is found to be approaching $\sim$2000 - apparently heavily dark matter dominated. Using N-Body simulations, we demonstrate that the observed luminosity, ellipticity, irregular morphology, velocity gradient, and the velocity dispersion can be well reproduced through processes associated with tidal mass loss, and in the absence of dark matter. These results highlight the considerable uncertainty that exists in  measurements of the dark matter content of Ursa Major II. The dynamics of the inner tidal tails, and tidal stream, causes the observed velocity dispersion of stars to be boosted to values of $>$5~km~s$^{-1}$. These dispersion boosts occur at each apocentre, and last throughout the time the galaxy is close to apocentre. The model need not be close to destruction to have a boosted velocity dispersion. We additionally note that the velocity dispersion at apocentre is periodically enhanced substantially (e.g $>$20~km~s$^{-1}$). This occurs most strongly when the model's trajectory is close to perpendicular with the Galaxy's disk at pericentre. This effect is responsible for raising the velocity dispersion of our model to (and beyond) the observed values in UMaII. We test an iterative rejection technique for removing unbound stars from samples of UMaII stars whose positions on the sky, and line-of-sight velocities, are provided. We find this technique is very effective at providing an accurate bound mass from this information, and only fails when the galaxy has a bound mass less than 10$\%$ of its initial mass. However when $<$2$\%$ mass remains bound, mass overestimation by $>$3 orders of magnitude are seen. Additionally we find that mass measurements are sensitive to measurement uncertainty in line-of-sight velocities. Measurement uncertainties of 1-4~km~s$^{-1}$ result in mass overestimates by a factor of $\sim$1.3-5.7.
\end{abstract}


\section{Introduction}
The number of galaxies that are considered satellites of the Milky Way
(MW) has grown substantially in the last decade as a result of careful
analysis of the SDSS catalogue \citep[e.g.][and many
more]{Willman2005,Bel2006,Zucker2006,Bel2007a,Walsh2007,Bel2010}.  The new members are faint
low surface brightness dwarf spheroidal galaxies (dSph), many of which
are less luminous than a globular cluster or even an open cluster.
Measurements of their internal velocity dispersion present suprisingly
high values in comparison to their luminosities
\citep[e.g.][and more]{Simon2007,Koch2009,Geha2009}.  Under the assumption that
these systems are in virial equilibrium, they would appear to be the
most dark matter dominated objects yet discovered, with mass-to-light
ratios greater than a thousand in some cases
\citep[e.g.][]{Fellhauer2008,Simon2007,Geha2009}.  

These new galaxies are an important discovery, as they close the gap
between observed numbers of dwarf galaxies and those predicted by
$\Lambda$CDM models - the so-called `missing satellite problem'
\citep[e.g.][]{Klypin1999,Moore1999b}.  While $\Lambda$CDM has proven highly
successful in explaining evolution of the large-scale structure of the
Universe \citep[e.g.\ Millennium II simulation of][]{Boylan-Kolchin2009}, it
additionally predicts that the dark matter halo of a MW-like galaxy 
should be surrounded by hundreds to thousands of small dark matter
halos \citep[e.g.\ Via Lactea INCITE simulation of][]{Kuhlen2008}.  Such
halos could host dwarf galaxies, if they can first capture, and then
maintain their baryonic components against a variety of complex,
non-linear, and poorly understood physical mechanisms. \citet{Koposov2009}
suggest that during the era of pre-reionisation, halos of mass less
than $v_{\rm circ} < 35$~km\,s$^{-1}$ would not have a sufficiently
strong potential to have gas-accretion.  At reionisation all the small
haloes will have their gas content ionised, halting their star
formation and stunting their growth.  Therefore, dwarfs with high
mass-to-light ratios are expected in $\Lambda$CDM when combined with a
baryonic physics recipe that is effective at removing gas from low
mass galaxies.  Some authors \citep{Tollerud2008} even claim that with
the discovery of the new dSphs, and allowing for more distant dwarfs
and regions of the sky not covered by the SDSS, the missing-satellite
problem may be solved.  

Other authors claim that new dSphs provide evidence against an origin
based in the $\Lambda$CDM paradigm.  Colour-magnitude diagrams reveal
that all the observed MW dwarfs have a population of ancient stars as
predicted within $\Lambda$CDM.  However, a number of MW dwarfs
demonstrate periods of quiescence followed by bursts of star formation
activity \citep[e.g.][]{deJong2008}.  Others appear to still form stars
today \citep{RyanWebber2008}.  These dwarfs show little evidence for the 
predicted halt in star formation during the reionisation
era.  Furthermore, the combined classical and new dwarfs around the MW 
appear to be aligned in a disc-like structure
\citep[e.g.][]{Metz2008,Metz2009}.  Such a disk-like
alignment of satellites is predicted to be highly unlikely within the
$\Lambda$CDM paradigm (\citealp{Pawlowski2012}).  The plane of this disk structure is close to
perpendicular to the plane of the MW, prompting the authors to suggest
that the MW dwarfs are in fact tidal dwarf galaxies.  These may have
been produced when another galaxy tidally interacted with the MW,
resulting in tidal dwarf galaxies spread along the orbit of the
interaction.  

A number of the new dSphs present possible indications of tidal
disruption \citep{Zucker2006b,Fellhauer2007,Munoz2008}, as indicated by irregular and
elongated morphologies \citep{Coleman2007,Sand2009}.  The measured
mass-to-light ratios of the new dSphs intrinsically rely on the
assumption of virial equilibrium -- an assumption that is violated
during the process of tidal disruption.  This, and the streaming of
unbound stars could in principle inflate the observed mass-to-light
ratio of a dSph signicantly beyond that of its true mass-to-light 
ratio (\citealp{Kroupa1997}; \citealp{Kuepper2011}; \citealp{Casas2012}). \citet{Kroupa2010} report that a simulation model galaxy
matching the properties of the Hercules dSph, presents an apparent
mass-to-light ratio of $\sim$200, although it contains no dark matter
at all. The presence of a high binary fraction could also boost the
apparent mass-to-light ratio of a dSph, although \citet{Minor2010}
suggest that this could only boost the real velocity dispersion by at
most $\sim$30$\%$ for dSphs with observed velocity dispersions of 4-10
km~s$^{-1}$. 

By virtue of its elongated shape, Ursa Major II (UMaII) has long been suspected to be
in the process of tidal disruption. This prompted \citet{Fellhauer2007} to
investigate if UMaII might be the possible progenitor of the Orphan
Stream \citep{Bel2007b}. The model of \cite{Fellhauer2007} was able to reproduce all the
observational data at the time, but only because the dSph was seen more or
less at the exact moment of final dissolution. The reactions of the community to
this kind of model were that this might be the case
for one dwarf, but cannot be a possible explanation for all dSph
galaxies of the MW (private comm. Fellhauer). Also, recent deep
photometry of UMaII has weakened the evidence for a direct connection
between UMaII and the Orphan Stream, as the stream appears misaligned
with UMaII's elongation \citep[see for example Fig.~10 of][]{Munoz2010}, although they were well aligned in \cite{Fellhauer2007}.
Furthermore, the surface brightness is now better constrained, and the
irregular morphology appears to continue until the very inner radii.
The dynamical properties of UMaII are also now better constrained in
\citet{Simon2007}.  i.e. UMaII has a velocity dispersion of $6.7 \pm
1.4$~km\,s$^{-1}$, and appears to have a velocity gradient across its
body, resulting in the West and East side of the galaxy differing in
radial velocity by $ 8.4 \pm 1.4$~km\,s$^{-1}$. A summary
of the key properties of UMaII is provided in Tab.~\ref{tabUMaII}. 

These new observations have prompted us to re-model UMaII, without assuming a connection between UMaII and the Orphan Stream. We attempt to reproduce the properties of UMaII, without relying on the assumption that the galaxy is on the verge of destruction to reproduce the high velocity dispersion. We consider three orbital trajectories, and test if our models reproduce the observed central surface brightness, ellipticity and
morphology, radial velocity, radial velocity gradients, and radial velocity dispersion. By comparison between the three models, we attempt to understand how the various properties of UMaII might be reproduced under the action of the MW tides alone, even in the absence of dark matter.

We describe the set-up and initial conditions of our simulations in Section 2, we present the results of our Fiducial Model in Section 3, and then compare this model with two other models so as to better understand the success of the Fiducial Model, in Section 4 we test the effects of low number statistics, and velocity measurement error on our results. Finally we summarise and conclude in Section 5. 

\begin{table}
\centering
\caption{Summary of key Ursa Major II properties. Table is split into photometric properties (upper section, taken from \citealp{Munoz2010}), and dynamical properties (lower section, taken from \citealp{Simon2007})}
\begin{tabular}{|c|c|}
\hline
  Right ascension (J2000) & 08 51 30.0\\
  Declination (J2000) & 63 07 48.0 \\
  Heliocentric Distance D$_\odot$&30$\pm$5 kpc\\
  V-band Absolute Magnitude M$_{\rm{V}}$ & -3.9 $\pm$ 0.5\\
  V-band Central Surface brightness $\mu_0^{\rm{V}}$ & $\sim$ 29 mag$_{\rm{V}}$ arcsec$^{-2}$\\
  Ellipticity $\epsilon$ & 0.5 $\pm$ 0.2\\
\hline
  Radial Velocity V$_{\rm{R}}\odot$ & -116.5 $\pm$ 1.9 km~s$^{-1}$\\
  Velocity gradient (E-W) V$_{\rm{diff}}$ & 8.4 $\pm$ 1.4 km~s$^{-1}$\\
  Velocity dispersion $\sigma$ & 6.7 $\pm$ 1.4 km~s$^{-1}$\\
\hline
\end{tabular}
\label{tabUMaII}
\end{table}

\section{Setup}

\subsection{Orbits}
To determine a possible orbit for UMaII, we perform test particle integrations in a Milky Way potential. This consists of a spherical, logarithmic halo of the form
\begin{equation}
\Phi_{\rm{halo}}(r)=\frac{v_o^2}{2}\ln(r^{2}+d^2)
\end{equation}

\begin{table}
\centering
\caption{A sample of possible orbits for UMaII that match the current right-ascension and declination, heliocentric distance and radial velocity of UMaII, a proper motion vector aligned with the UMaII elongation, an orbit that remains within the MW halo (r$_{\rm{apo}}$ $\le$ 250 kpc), and a radial velocity gradient greater than 1.0 km s$^{-1}$ per degree of right ascension. Columns are (i) model label: Fiducial Model (FM), Comparison Models 1 \& 2 (CM1 \& CM2), (ii) current proper motion in right ascension $\mu_\alpha$, (iii) current proper motion in declination $\mu_\delta$, (iv) pericentre distance r$_{\rm{peri}}$, (v) apocentre distance r$_{\rm{apo}}$, (vi) orbital eccentricity $e$, and (vii) orbital period (P)}
\begin{tabular}{c|c|c|c|c|c|c}
\hline
  Model & $\mu_\alpha$ & $\mu_\delta$ & r$_{\rm{peri}}$ & r$_{\rm{apo}}$ & $e$ & P\\
   & (mas yr$^{-1}$) & (mas yr$^{-1}$) & (kpc) & (kpc) & & (Gyr) \\
 \hline
  FM & -0.3 & -1.4 & 2.4 & 36.5 & 0.87 & 0.43 \\
  CM1 & 2.0 & -3.2 & 32.5 & 192.6 & 0.71 & 2.60 \\
  CM2 & 0.4 & -1.65 & 3.4 & 37.0 & 0.83 & 0.45 \\
\hline
\end{tabular}
\label{orbitpars}
\end{table}

\noindent
with $v_0$=186 km~s$^{-1}$ and d=12 kpc. The disk is represented by a Miyamoto-Nagai potential (\citealp{Miyamoto1975}):
\begin{equation}
\Phi_{\rm{disk}}(R,z)=\frac{-G M_{\rm{d}}}{\sqrt{R^2+(b+\sqrt{z^2+c^2})^2}}
\end{equation}
\noindent
with $M_{\rm{d}}$=10$^{11}$ M$_\odot$,$b=6.5$~kpc and $c=0.26$~kpc. Finally the bulge is modelled as a Hernquist potential (\citealp{Hernquist1990}):

\begin{equation}
\Phi_{\rm{bulge}}(r)=\frac{-G M_{\rm{b}}}{r+a}
\end{equation}
\noindent
using $M_{\rm{b}}=3.4 \times 10^{10}$~M$_\odot$ and $a=0.7$~kpc. This choice of parameter values for the Milky Way's halo, disk and bulge were first used in \cite{Johnston1995}, and have since been frequently used to approximate the potential of the Milky Way (e.g. \citealp{Dinescu1999}; \citealp{Sakamoto2003}; \citealp{Fellhauer2008}; \citealp{Assmann2011}). The superposition of these three potentials provides a reasonable representation of the Milky Way potential field with a circular velocity at the solar radius (8~kpc from the Galactic centre) of $\sim$220~km~s$^{-1}$ (\citealp{BT2008}). We assume the Sun moves with respect to the LSR with a velocity ($V_{\rm{x}}$, $V_{\rm{y}}$, $V_{\rm{z}}$) = (10.0, 5.3, 7.2)~km~s$^{-1}$ (\citealp{Dehnen1998}).

Initially we use trial and error to find a suitable orbit which can match a number of requirements of the observational data. These requirements include a final position in the sky that matches the current right-ascension and declination, heliocentric distance and radial velocity of UMaII (see Tab. \ref{tabUMaII}). We assume that the motion vector on the sky at the current epoch matches the direction of the elongation of UMaII. Numerous orbits are found that can meet these constraints. We further constrain the orbit by requiring that the resulting test particle orbits cannot have an apocentre of beyond the extent of the MW halo (assumed to be $\sim$250~kpc).

In Tab. \ref{orbitpars}, we present the key parameters of three orbits that obey these constraints. Column 1 indicates the model, columns 2 \& 3 are the proper motion on the sky, columns 4 \& 5 are pericentre and apocenter distance respectively, column 6 is the eccentricity, and column 7 is the orbital period. Our best match to the observations is the Fiducial Model. This has the most plunging (i.e. smallest r$_{\rm{peri}}$) of all the orbits and, with an apocentre of 36.5~kpc, the model is found near apocentre when it reaches UMaII's current position in the sky. To better understand the success of our Fiducial Model, we compare it to two other models -- Comparison Model 1 (Model CM1) and Comparison Model 2 (Model CM2). Model CM1 has an orbit that is a little less eccentric than that of the Fiducial Model, and has a significantly larger pericentre and apocentre distance, resulting in weaker tidal forces throughout the orbit, and a substantially longer orbital period. Model CM2 has an orbit similar to the Fiducial Model but reversed, and slightly less eccentric and less plunging.

\subsection{Live progenitor models for UMaII}

Once we have chosen an orbit that matches the orbital constraints, we compute the orbit backward 10 Gyr. We then insert a live progenitor model for UMaII, and evolve the simulation forward to the current position of UMaII using the particle mesh-code {\sc{Superbox}} (\citealp{Fellhauer2000}). As we have worked backwards from UMaII's current position, and velocity, the models automatically match the observed galaxy's position, and line-of-sight velocity, simply as a result of this approach.

{\sc{Superbox}} is an efficient and fast particle-mesh code, whose use of a fast, low storage FFT-algorithm enables multi-million particle simulations to be performed on desk-top computers.  {\sc{Superbox}} uses high resolution sub-grids, that follow the trajectories of a galaxy, making it highly suited for studies of dwarf galaxy tidal disruption.

For our live progenitor model, we choose a Plummer model. The spatial and velocity distribution of particles within a Plummer model is controlled chiefly by two parameters - the total mass M$_{\rm{pl}}$, and scalelength R$_{\rm{pl}}$. Increasing M$_{\rm{pl}}$ or decreasing R$_{\rm{pl}}$ also increases how robust the Plummer model is to tidal disruption. In practice, we vary both parameters in order to best match the observed luminosity, central surface brightness, morphology and ellipticity of UMaII, while simultaneously attempting to match the observed dynamics -- the line-of-sight velocity gradient, and the velocity dispersion. Our progenitor models consist of 1 million N-body particles. We choose a {\sc{Superbox}} subgrid with a spatial resolution of $\sim$one-tenth that of R$_{\rm{pl}}$, for the high resolution grid. This ensures we resolve the Plummer sphere sufficiently to ensure stability in isolation. We note that our spatial resolution suppresses the formation of binaries, and other two-body effects. As such, all mass loss occurs as a result of gravitational tides. We evolve the progenitor model along its orbit for 10 Gyr, under the influence of the MW tides, until it reaches its current location in right-ascension and declination. At this point we measure its properties and compare to the observed properties.

Our initial conditions are purely baryonic (i.e. do not contain a component of dark matter), as we are interested in demonstrating how tidal mass loss alone can reproduce many of the observed properties of UMaII -- including the high mass-to-light ratio. However our initial model could also be regarded as having been previously dark matter dominated, but then tidally stripped of its dark matter content prior to the time at which our simulations begin.

\begin{figure}
  \centering \epsfxsize=8.5cm \epsfysize=14.0cm \epsffile{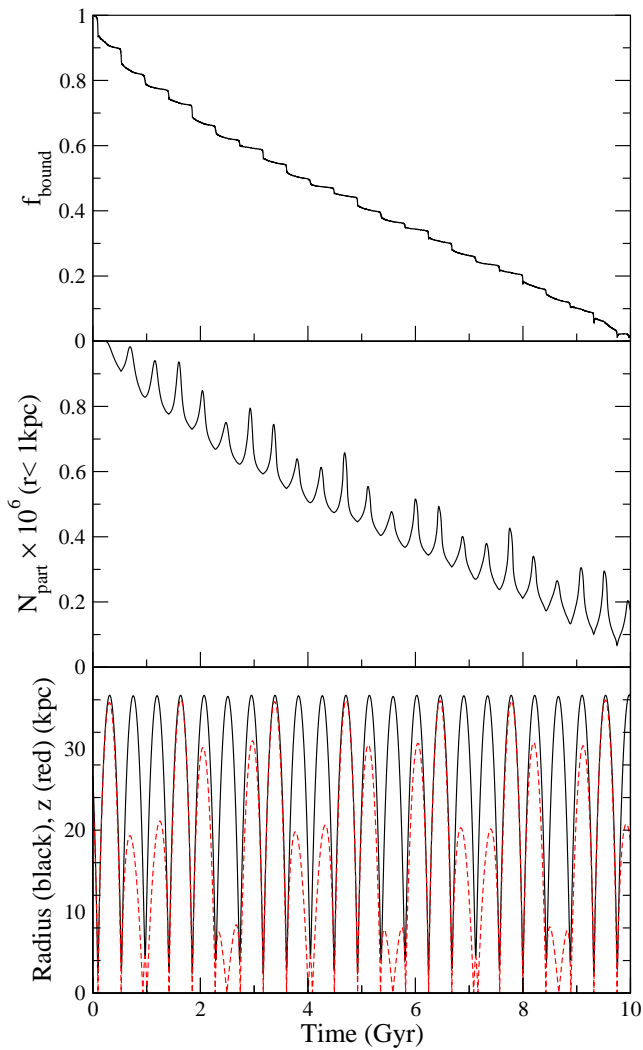}
  \caption{Time evolution of the Fiducial Model's bound mass fraction (upper panel), (centre panel) the number of particles within 1~kpc of the centre of density, and (lower panel) the orbital radius (solid, black line) and z-position (dashed, red line). At each pericentre a roughly equal quantity of mass is unbound. At each apocentre, contraction of the tidal streams causes a temporary peak in the number of stars near the bound core.}
\label{massloss}
\end{figure}

\section{Results}
\subsection{The Fiducial Model}
\subsubsection{Evolution from the progenitor until now}
The Fiducial Model has an initial progenitor with a total Plummer mass M$_{\rm{pl}}$ of 7.57$\times10^5$~M$_\odot$, a Plummer scalelength R$_{\rm{pl}}$=11.7~pc, and is radially truncated at 500.0~pc. The orbital properties can be found in Tab. \ref{orbitpars}. The plane of the orbit is inclined at approximately 90$^\circ$ to the plane of the Milky Way disk.

The upper panel of Fig. \ref{massloss} shows the time-evolution of the bound mass fraction of the Fiducial Model. To measure if a particle is bound, we use the `snowballing' method described in \cite{Smith2013}. The lower panel shows the corresponding orbital radius (solid black line), and the z-position (dashed red line) of the centre of density of the model. The radial evolution (solid black line) shows successive apocentre passes (visible as peaks), and pericentre passes (visible as troughs). A complete orbit has $\sim$430~Myr duration. The z-evolution (dashed red line) demonstrates that orbital precession occurs -- once every 3-4 orbits the z-position becomes equal to the apocentre distance. When this occurs, then the previous pericentre pass occurs along a trajectory that is almost perpendicular to the plane of the MW.

Comparing the upper and lower panel we see that, with each pericentre pass, a small fraction ($\sim$4$\%$) of the mass is tidally stripped resulting in a repeated step-like decrease in the bound fraction. The mass loss at each pericentre pass is roughly equal. After 10~Gyr evolution $>$95$\%$ of the mass has been stripped.

\begin{figure*}
  \centering \epsfxsize=18.5cm  \epsffile{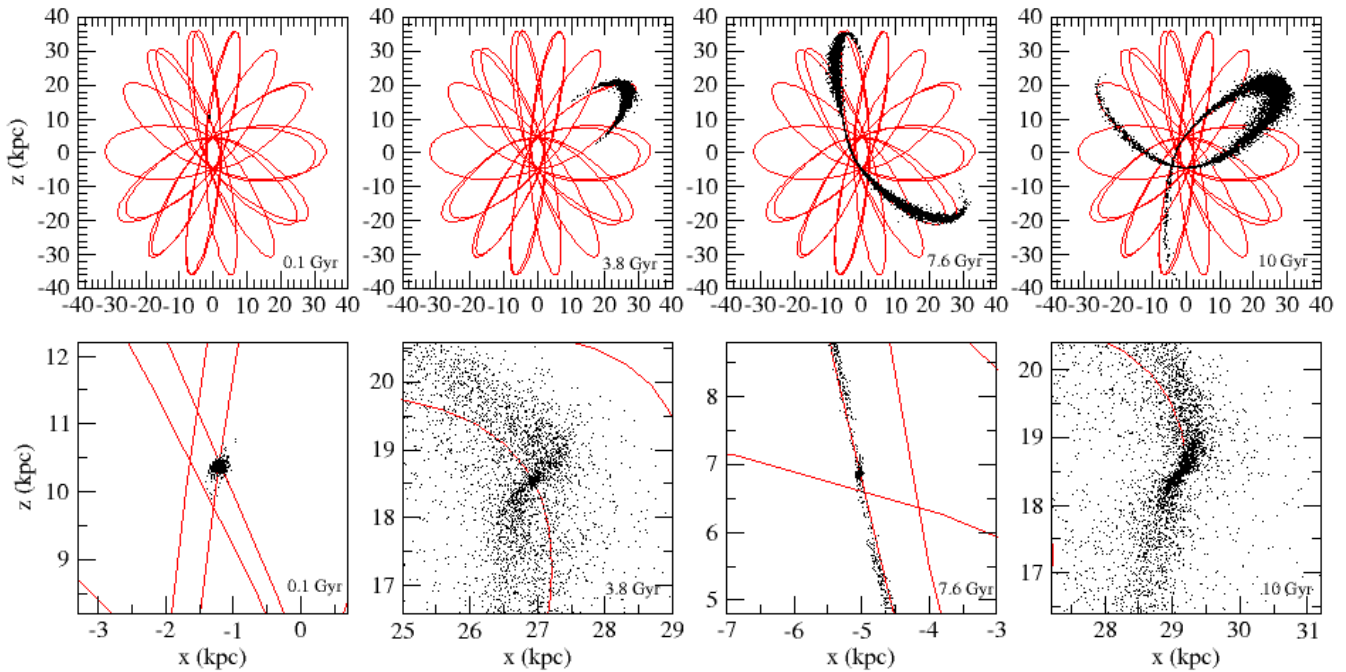}
  \caption{x-z plots of the fiducial model's particle distribution at varying times throughout the orbit (time shown in lower-right corner). Upper rows shows the full orbital trajectory (80 by 80 kpc), while the lower row is centred on the bound remnant (4 by 4 kpc). The red curve is the orbital trajectory of the galaxy in the x-z plane over the whole 10~Gyr. }
\label{xzsnaps}
\end{figure*}

Fig. \ref{xzsnaps} shows snapshots of the particle distribution of the Fiducial Model throughout its evolution (time of snapshot in indicated in the lower right corner). The upper row is the whole 80 by 80 kpc region containing the orbit. The red curve indicates the orbital trajectory over the complete 10~Gyr simulation, and shows the orbital precession about the plane of disk (where the plane of the disk is in the x-y plane). The lower panel is a 4 by 4 kpc region centred on the bound remnant of the Fiducial Model. 

At t=0.1~Gyr (left column), the model has just passed pericentre for the first time, and particles that have been unbound are starting to leave the bound core. By t=3.8~Gyr, the model has passed pericentre on 9 previous occasions, and is currently close to apocentre. The upper panel reveals that the tidal streams have grown in length, and follow the orbital trajectory. The lower panel shows that from a small bound core, stars are exiting at the Lagrangian points (\citealp{Kuepper2010}), resulting in `inner tidal tails' that roughly point towards the cente of the MW (\citealp{Klimentowski2009}). $\sim$1~kpc from the bound core, these inner tidal tails convert into tidal streams that quickly orientate themselves along the model's trajectory. At t=7.6~Gyr, the model has just passed pericentre. The upper and lower panel reveal that the tidal streams continue to grow in length, as increasing amounts of mass are liberated from a reduced bound core. Near pericentre, the stars are most stretched out along the orbit. At t=10~Gyr (the right column), the Fiducial Model is located in the current position of UMaII in the sky, and is just past the apocentre of the orbit. 

Comparing the lower panel of t=7.6~Gyr and t=10~Gyr, it can be seen that the unbound stars in tidal streams bunch up closely around the bound remnant of the model when it is at apocentre. We count $\sim$5 times as many unbound stars in the lower panel of t=10~Gyr (at apocentre) as at t=7.6~Gyr (at pericentre). This occurs as the tidal tails of stripped stars tend to contract about the bound core when the core is near apocentre. In the centre panel of Fig. \ref{massloss}, we plot time evolution of the total number of particles within 1~kpc of the model's centre of density. Clearly the total number of particles decreases as the bound mass of the remnant is stripped away. However a peak in the number of particles also occurs at each apocentre as a result of the contraction of the tidal tails. The peaks are slightly larger when distance from the plane $z\sim$ orbital radius $R$. Similar behaviour is seen in the velocity dispersion of the stars, as discussed in detail in Section \ref{veldispboosts}.

\subsubsection{Final properties of the Fiducial Model after 10~Gyr evolution}
\label{model1final}

\label{Fiducialevol}
\begin{figure}
\begin{center}$
\begin{array}{c}
\includegraphics[width=6.8cm]{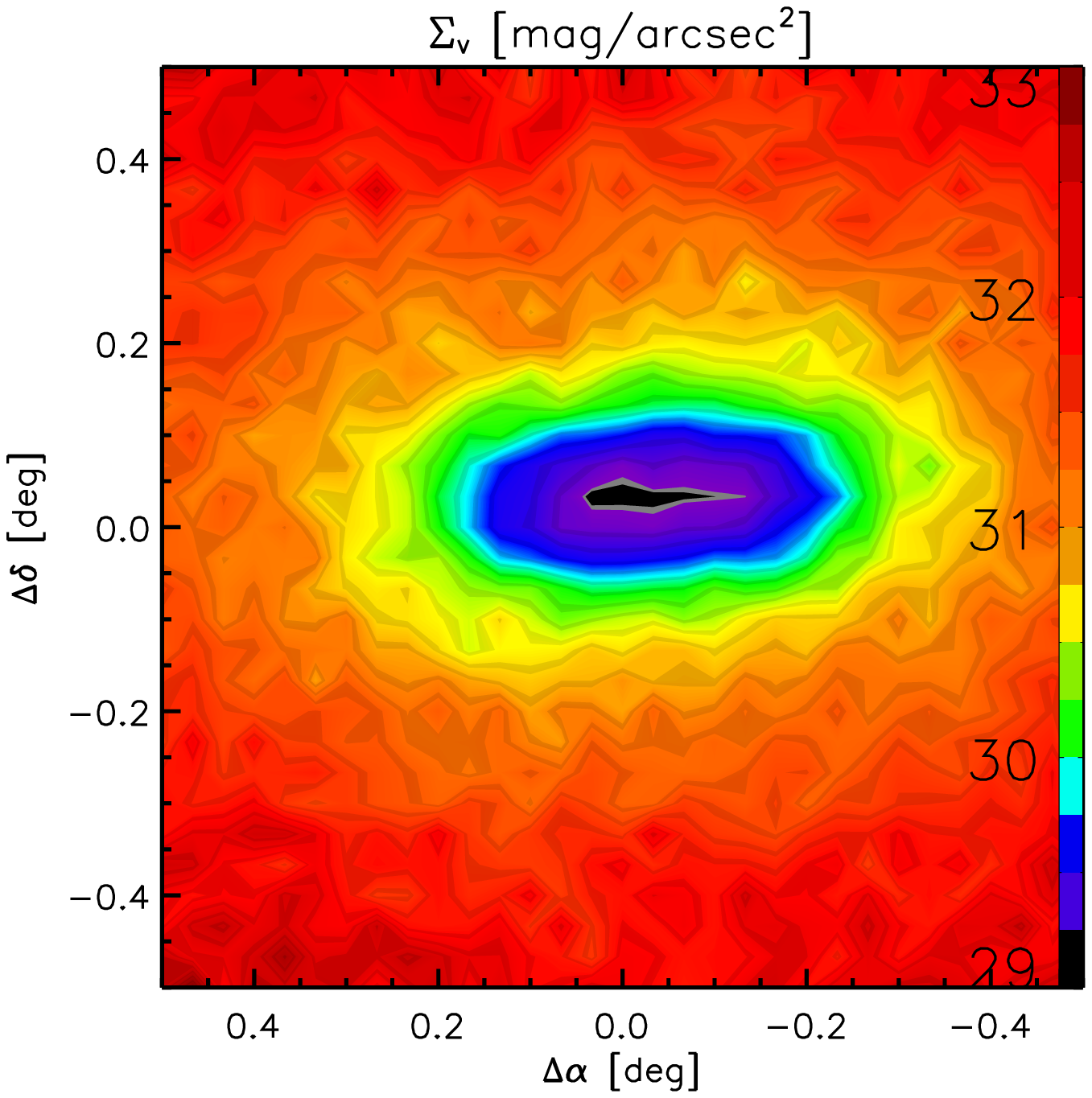} \\  
\includegraphics[width=6.8cm]{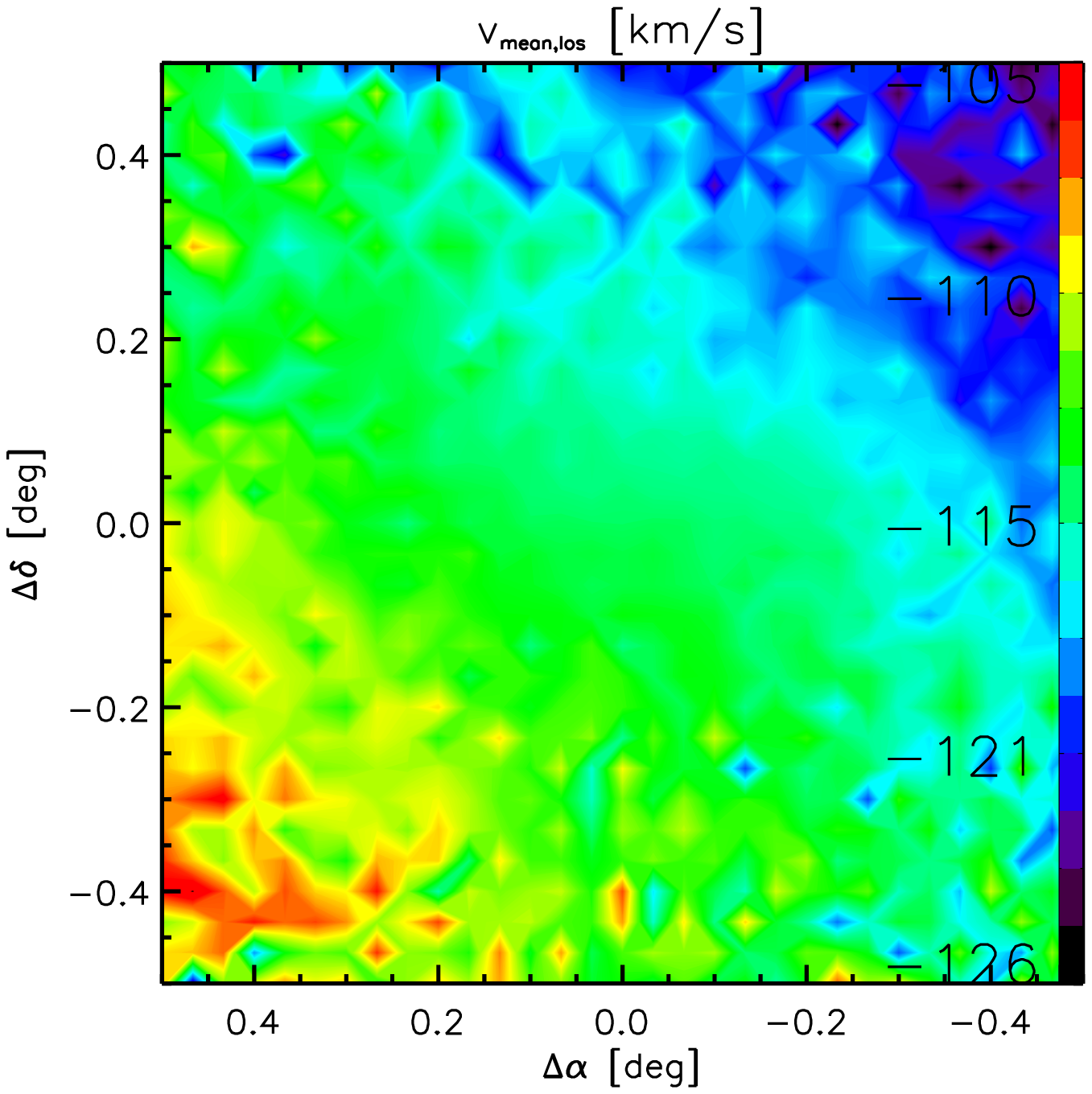} \\  
\includegraphics[width=6.8cm]{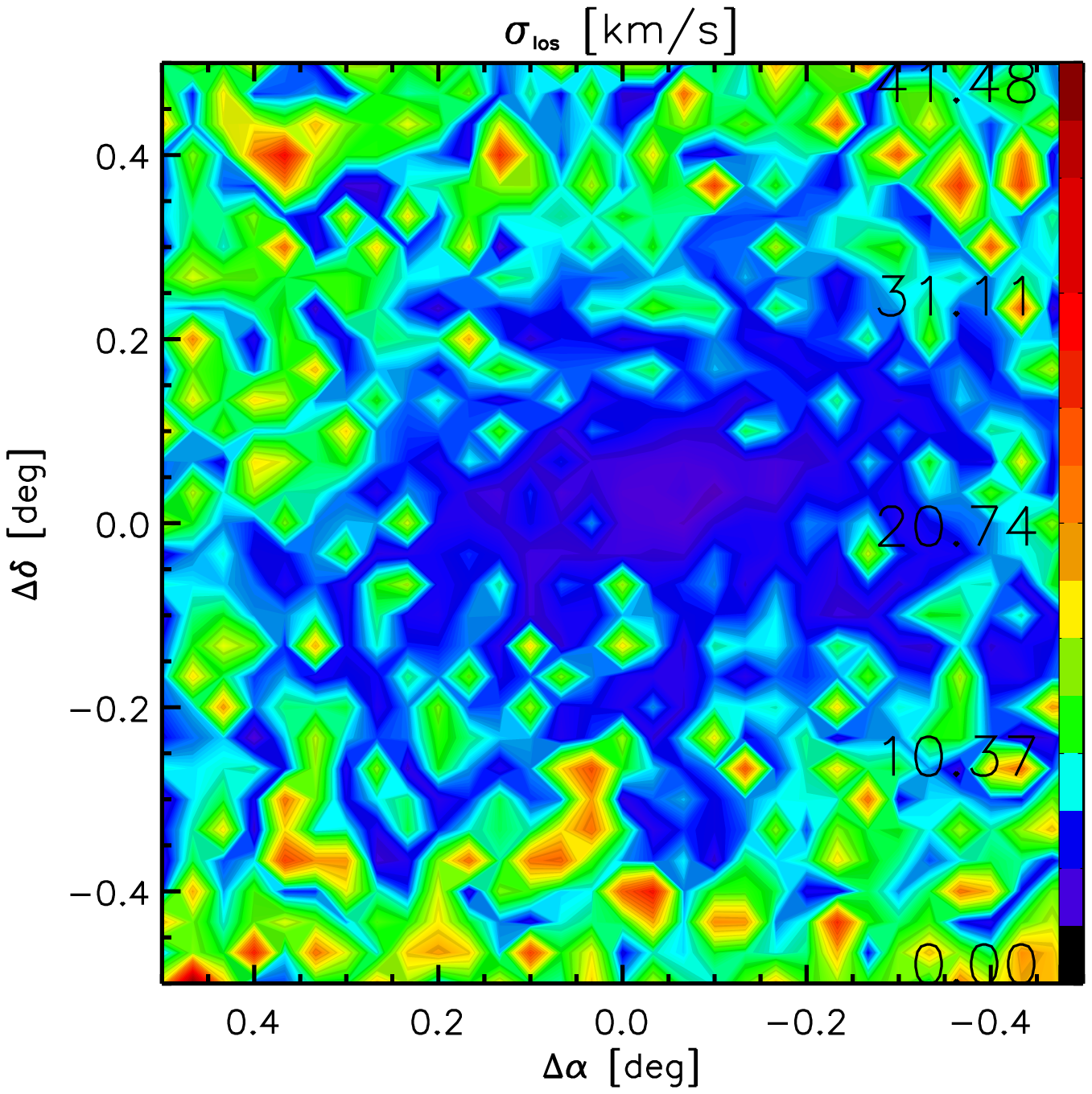} \\  
\end{array}$
\end{center}
\caption{Final properties of the Fiducial Model. After 10~Gyr of tidal mass-loss while orbiting in the MW potential, the final remnant has arrived at the observed location of UMaII. The upper panel is surface brightness assuming a stellar mass-to-light ratio of 1.7. The middle panel is the average velocity down our line-of-sight, while the lower panel is the velocity dispersion down our line-of-sight. Images have 20~pc resolution, matching that of the observed stellar number-density map (Mu\~noz et al. 2012). Colour bars indicate surface brightness contours in mag$_{\rm{V}}$ arcsec$^{-2}$ (top panel) and velocity contours in km~s$^{-1}$ (centre and lower panel).}
\label{finalgal1}
\end{figure}

After 10~Gyr of evolution, our model galaxy arrives at the current position of UMaII in the sky, having evolved substantially under the influence of the MW tides. Fig. \ref{finalgal1} shows how the Fiducial Model appears in surface brightness (top panel), line-of-sight velocity (middle planel), and velocity dispersion (lower panel). 

The upper panel of Fig. \ref{finalgal1} is surface brightness of the Fiducial Model in mag$_{\rm{V}}$ arcsec$^{-2}$ (as indicated by the colour bar on the right). To produce this plot, we assume a stellar mass-to-light ratio of 1.7. This results in our galaxy having a V-band absolute magnitude of M$_{\rm{V}}$=-4.4, consistent with the observed value of M$_{\rm{V}}$=-3.9$\pm$0.5. 

A visual comparison can be made with Fig. 11 of \cite{Munoz2010}, where the observed isodensity contour maps are shown. We note that, in their figure, these contours are not literally contours in surface brightness. Rather they are contours joining regions where star number densities are a specified number of the standard deviation of star number densities measured in a background region (from regions outside the main body of UMaII). \cite{Munoz2010} note that their three sigma contour (the outer-most contour in their figure) is roughly equivalent to a V-band surface brightness of 32.6 mag$_{\rm{V}}$~arcsec$^{-2}$. Assuming this value as exact, we convert the remaining contours into units of mag$_{\rm{V}}$~arcsec$^{-2}$. In this way, we find the inner-most contour is equivalent to 29.2 mag$_{\rm{V}}$~arcsec$^{-2}$. However we note that there is some uncertainty in the true surface brightness due to our conversion from star counts per unit area to surface brightness.

Our model galaxy has central surface brightness of 29.2 mag$_{\rm{V}}$~arcsec$^{-2}$, in excellent agreement with the observed value we derived from the star counts. The direction of elongation of the Fiducial Model also matches the observed direction, and lies along the orbital trajectory of the Fiducial Model. The most bound particles are found close to the centre of the galaxy, and form the highest surface brightness regions. The model was initially spherical but, as the model is very close to complete disruption (e.g. see lower panel of Fig. \ref{massloss}), there are indications of tidal distortion at all radii within the galaxy. As such the ellipticity rises from 0.4 to 0.6 with increasing radius from the galaxy core, in good agreement with the observed values, and the trend with radius. Surrounding the bound core are streams of unbound particles, causing the surface brightness to fall to low values ($>$32 mag$_{\rm{V}}$~arcsec$^{-2}$) at 0.2-0.4 degrees from the bound core.

\begin{figure}
  \centering \epsfxsize=8.5cm  \epsffile{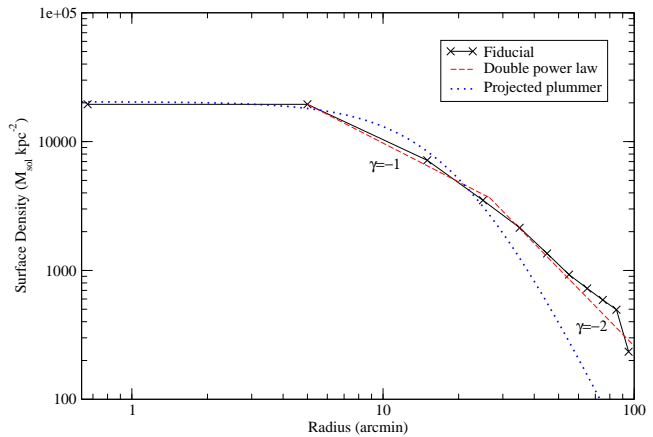}
  \caption{The radial surface density profile of the Fiducial Model (black line with crosses). The model is poorly fit by a projected Plummer profile (dotted blue line). However the intermediate and outer radii are well fitted by a double power law profile, with a slope of $\gamma$=-1.0 and -2.0 respectively, with the break between the profiles occuring at $R\sim$20~arcmin. This is in a reasonable agreement with the observed profile. However at small radii ($R<$5~arcmin) the model approaches $\gamma$=0, at odds with the observed profile. This is likely due to the cored density profile of the progenitor model.}
\label{surfrad}
\end{figure}

The observed radial surface density profile of UMaII is well matched by a double power law with an inner slope of $\gamma$=-0.96, out to a radius of $\sim$20~arcmin, and an outer slope of $\gamma$=-2.40 (see Fig. 7 of \citealp{Munoz2010}). This form of profile is at odds with the `projected-Plummer' form ($\Sigma(R)=\Sigma_0/\left[1+(R/R_{\rm{PL}})^2\right]^2$) seen in simulations of dSphs, embedded in dark matter halos, that have undergone significant tidal mass loss on eccentric orbits within a MW potential (\citealp{Penarrubia2009}). The radial surface density profile of the Fiducial Model is shown in Fig. \ref{surfrad}. We compare our results (solid black line with crosses) with a double power law fit (dashed red line). We also compare to a projected-Plummer form (dotted line), whose best fit gives $R_{\rm{PL}}=20$~arcmin. Like the observed profile, our results are poorly fit by the projected-Plummer form, which falls off too quickly with increasing radius. At intermediate and large radius, our results are well fitted by a double power law. At intermediate radius ($R$=5-20~arcmin) the slope is $\gamma$=-1.0, in good agreement with the observed value ($\gamma$=-0.96). At large radius ($R$$>$20~arcmin) the slope is $\gamma$=-2.0, which is in reasonable agreement (although somewhat flatter) than the observed value ($\gamma$=-2.4). However at very small radii ($R<10$~arcmin) there is clear disagreement between the model (which tends to $\gamma=0$), and the observed profile (which maintains $\gamma$=-0.96 as $R\rightarrow$0). This is likely because our progenitor models are Plummer distributions, which have cored profiles. This perhaps suggests that a progenitor model must contain a central cusp at small radius ($R<50$~pc) to better match the inner-most part of the observed profile.

In the middle panel of Fig. \ref{finalgal1}, we show the average velocity of stars that would be observed in the Fiducial Model. This is  centred at roughly -115~km~s$^{-1}$. There is a clear velocity gradient in both right-ascension, and also in declination. We measure a velocity gradient in right ascension of 8.4~km~s$^{-1}$ per degree, across the body of the model in the direction of the right ascension. This is in excellent agreement with the observed value of 8.4$\pm$1.4~km~s$^{-1}$ per degree. 

In the lower panel of Fig. \ref{finalgal1}, we show the velocity dispersion of stars that would be observed in the Fiducial Model. Close to the core of the galaxy, the velocity dispersion is lowest, as this is where the bound stars are found. Meanwhile at $\sim$0.5 degrees from the core, the velocity dispersion can become very large ($\sim$30-40~km~s$^{-1}$) in some pixels, due to the streaming motions of unbound stars.

The dynamical properties of stars in UMaII, such as the line-of-sight velocity, velocity gradient, and velocity dispersion, were meaured in \cite{Simon2007}. However the stars of their sample were located in two strips, that meet to form an `L-shaped' region  of sky (see Fig. 2 of \citealp{Simon2007}). In order to fairly compare the dynamics of the models to that of the observations, we measure the radial velocity, velocity gradient, and velocity dispersion in the same `L-shaped' region in right ascension-declination space, and notate this as $\sigma_{\rm{L}}$. As a result we avoid the biggest peaks of the velocity dispersion seen in the lower panel of Fig. \ref{finalgal1}. For example, for the Fiducial Model we find $\sigma_{\rm{L}}$=6.29~km~s$^{-1}$. This is in good agreement with the observed velocity dispersion of 6.7$\pm$1.4~km~s$^{-1}$.

In summary, the Fiducial Model well reproduces the observed morphological properties of UMaII, including luminosity, central surface brightness, appearance, and eccentricity. The Fiducial Model also reproduces the observed dynamics, including the velocity gradient in right-ascension, and the observed velocity dispersion.

\begin{figure}
\begin{center}$
\begin{array}{c}
\includegraphics[width=7.cm]{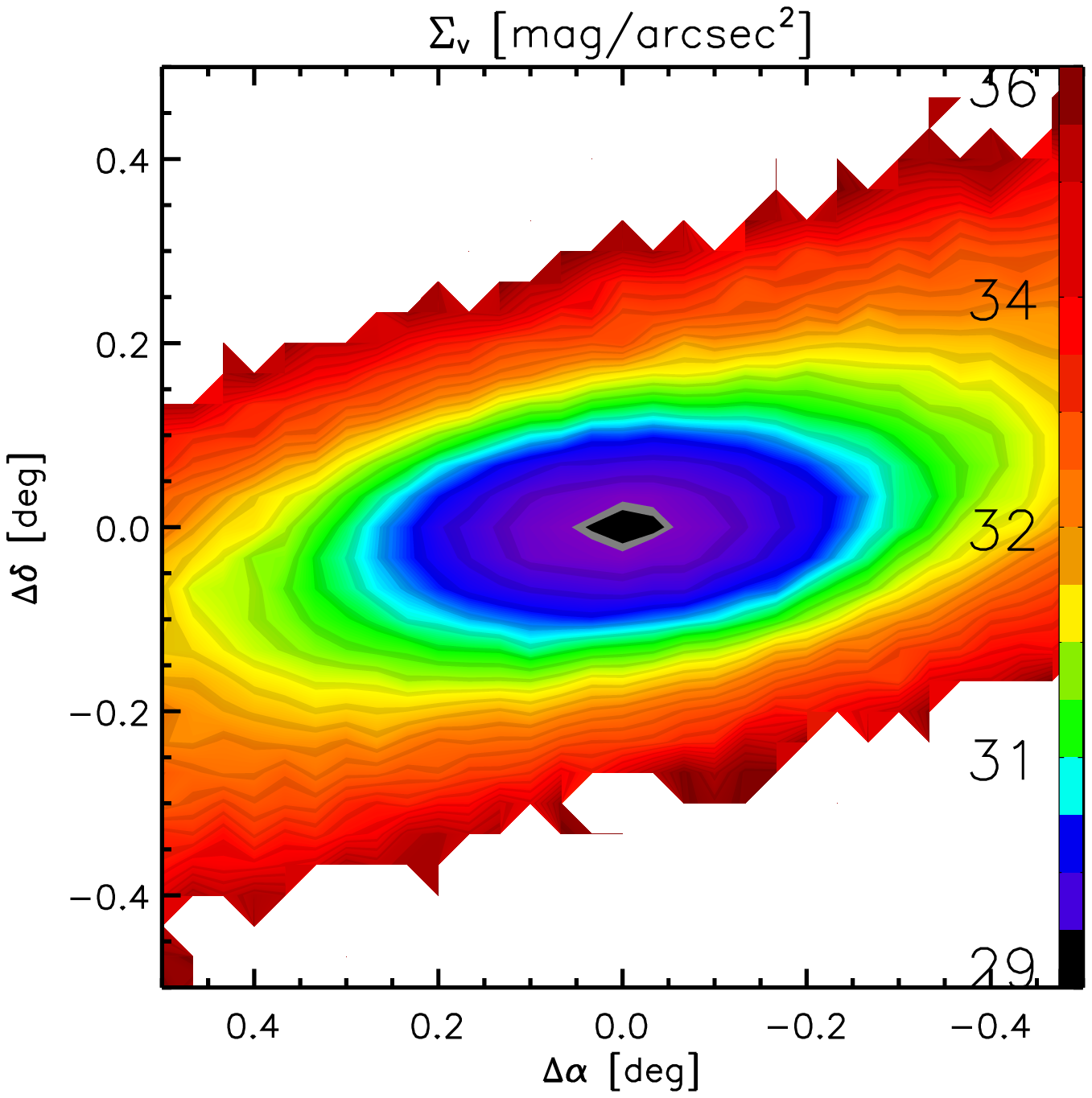} \\  
\includegraphics[width=7.cm]{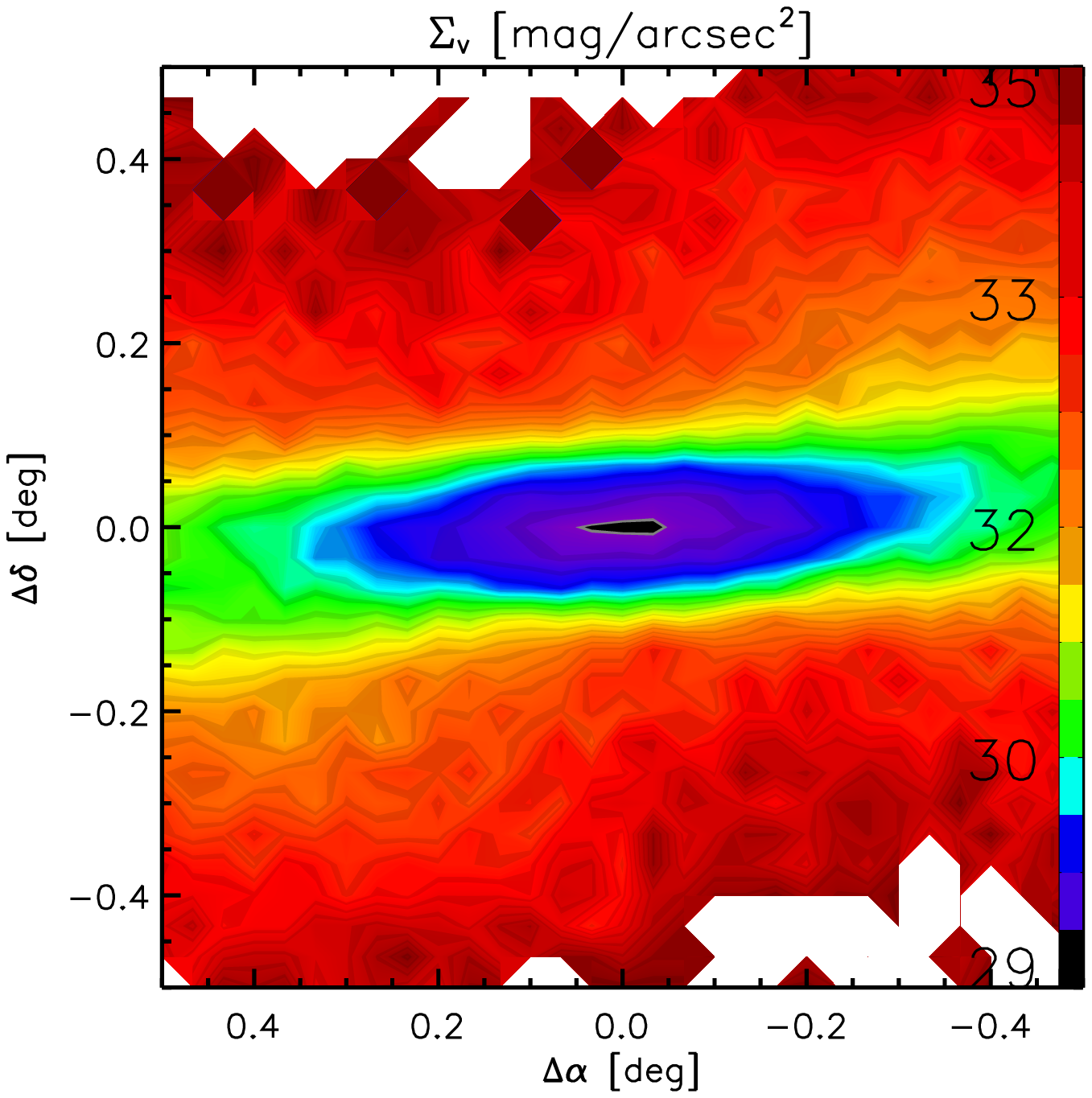} \\  
\end{array}$
\end{center}
\caption{Surface brightness of Model CM2 (upper) and CM3 (lower). Surface brightness is displayed in the colour bars in units of mag$_{\rm{V}}$ arcsec$^{-2}$. A stellar mass-to-light ratio of 2.0 and 2.2 is assumed for Model CM2 and CM3 respectively.}
\label{finalgal2}
\end{figure}

\subsection{How does the Fiducial Model match the observed properties of UMaII through tides alone?}
In order to better understand how the Fiducial Model is able to reproduce the observed properties of UMaII, we additionally run two other comparison models -- Comparison Model 1 (Model CM1), and Comparison Model 2 (Model CM2). The orbital properties of these models are given in Tab. \ref{orbitpars}. By comparing between the Fiducial Model and these comparison models, we can gain deeper insight into the mechanisms at work, and better understand how unique a solution the Fiducial model is (i.e. how easy it is to reproduce the observations of UMaII or, indeed, other dSphs).

\subsubsection{Photometric properties and general morphology}

As with the Fiducial model, we vary both the initial mass and scalelength of both comparison models, using a trial and error approach, in order to try and match the final luminosity, central surface brightness, and general morphology UMaII following 10~Gyr of orbital evolution. In Fig. \ref{finalgal2} we present the surface brightness maps of Model CM1 (top panel) and Model CM2 (lower panel) (n.b. the surface brightness map of the Fiducial Model is in the upper panel of Fig. \ref{finalgal1}). All three models are elongated along the orbital trajectory, matching the direction of elongation in UMaII. The following photometric properties of UMaII are taken from \cite{Munoz2010}. While UMaII has a luminosity of -3.8~M$_V$, the Fiducial model, Model CM1 and Model CM2 have luminosity -4.4, -3.5 and -4.0 respectively. Thus all three models are consistent with the observed luminosity within error bars. UMaII has an approximate central surface brightness of 29.1~mag$_{\rm{V}}$~arcsec$^{-2}$. All three models closely match the observed value with the Fiducial model, Model CM1 and Model CM2 having a central surface brightness of 29.2, 29.1 and 29.1~mag$_{\rm{V}}$~arcsec$^{-2}$ respectively. UMaII has a measured ellipticity of $\epsilon=$0.5$\pm$0.2, and presents an irregular morphology from core to outskirts. Our Fiducial model and Model CM1 match the observed ellipticity well with $\epsilon=$0.4-0.6, and 0.3-0.5 respectively. Model CM2 is more flattened with $\epsilon=$0.2-0.3, but this is not highly significant given the considerable uncertainties in producing the observed star count contours (e.g. see Fig. 12 in \citealp{Munoz2010}). All three models present a distorted morphology at all radii due to all models being close to total tidal disruption.

\begin{table}
\centering
\caption{Parameters of the initial conditions of the three models. Columns are (i) Model label; Fiducial Model (FM), Comparison Models 1 \& 2 (CM1 \& CM2), (ii) Plummer mass M$_{\rm{pl}}$, (iii) Plummer scalelength R$_{\rm{pl}}$, (iv) Cut-off radius R$_{\rm{CO}}$ }
\begin{tabular}{c|c|c|c|}
\hline
  Orbit & M$_{\rm{pl}}$(M$_\odot$) & R$_{\rm{pl}}$(pc) & R$_{\rm{CO}}$(pc) \\
 \hline
  FM & 7.57$\times$10$^5$ & 11.7 & 500.0\\
  CM1 & 2.00$\times$10$^5$& 61.9 & 500.0\\
  CM2 & 7.57$\times$10$^5$& 14.3 & 500.0\\
\hline
\end{tabular}
\label{icspars}
\end{table}

In summary, for all three models (and with differing orbits), we find it relatively easy to reproduce the observed morphology of UMaII in terms of luminosity, central surface brightness, ellipticity, and general morphology. This is achieved simply by finding the correct pairing of the Plummer mass and Plummer scalelength (shown in Tab. \ref{icspars}). For orbits that suffer stronger tides, a progenitor that is more robust to tidal disruption is required to survive for 10~Gyr. For example the Fiducial Model has a pericentre of only 2.4 kpc, in comparison to 32.5 kpc for Model CM1. Thus the Fiducial Model progenitor is both more massive (by a factor $\sim$4) and more concentrated (by a factor $\sim$5), in order to be sufficiently robust. However, despite the substantial differences in the progenitor and orbit between the Fiducial Model and Model CM1, the final morphological properties are similar (e.g. compare the upper panels of Figures \ref{finalgal1} and \ref{finalgal2}). We note that the radial surface density profile of our Fiducial Model matches the observed profile reasonably well, except at the inner-most radii. To better match the surface brightness profile at small radii, we would require a progenitor model with a more cuspy profile.

\begin{figure}
  \centering \epsfxsize=8.5cm \epsffile{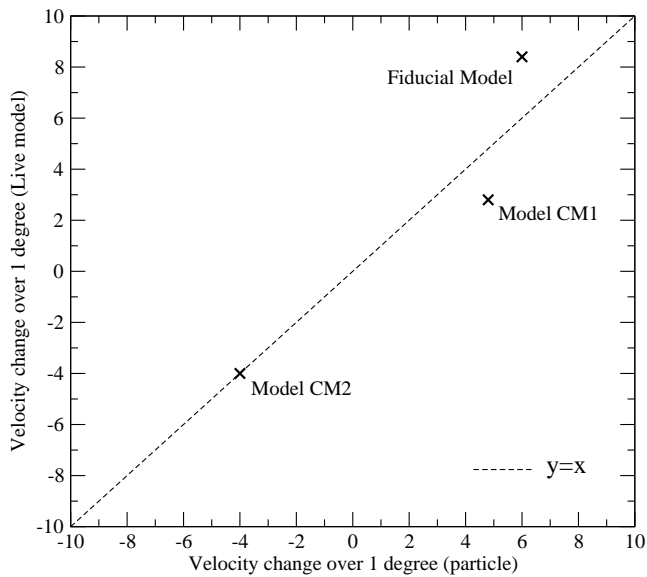}
  \caption{The difference in velocity across one degree of right-ascension at UMaII's position in the sky for; (x-axis) a single particle, (y-axis) the live N-body model galaxy on the same orbit. The diagonal dashed line indicates y=x. To first order the velocity gradients in all three models occur due to orbital acceleration. However the remaining bound mass of the remnant and streaming motions result in second order deviations away from the dashed line.}
\label{velgradcomp}
\end{figure}

\subsubsection{Velocity gradient from East-to-West}
In all three models (the Fiducial Model, Model CM1 and Model CM2), we find a velocity gradient with right-ascension across the galaxy body. We emphasise that our models are non-rotating, and therefore the velocity gradient is NOT a signature of rotation (despite its apparent similarity to that which would be produced by rotation).

We measure the amount by which the line-of-sight velocity changes over 1 degree of right-ascension in all three models when they arrive at UMaII's position in the sky. These are plotted on the y-axis of Fig. \ref{velgradcomp}. We also place a single particle on the orbit of each of the models, and measure the change in line-of-sight velocity over 1 degree of right-ascension. These are plotted on the x-axis of Fig. \ref{velgradcomp}. If the full, live model gradient agreed exactly with the single particle gradient it would fall on the dashed, diagonal line, and this would signify that the velocity gradient was simply the result of the change in orbital velocity of particles as they move along their orbital trajectory. For example, particles climbing out of the MW potential well will deaccelerate, while those falling in accelerate. We find that to first order, this is indeed the primary mechanism creating the velocity gradient. For example, the negative velocity gradient of Model CM2 could be predicted from the orbit of a single particle alone. 

However second order effects arise, as the particles within the live model are not all freely streaming in the MW potential, and this causes the results to scatter vertically about the dashed line. For example, the gravitational influence of the bound remnant of each model, and the releasing of unbound stars through inner tidal tails towards the Galactic centre (for example, see lower panel at t=3.8~Gyr) could result in some of the scatter in Fig. \ref{velgradcomp}.

In summary, we find that the observed presence of a gradient in the line-of-sight velocity across the body of the remnant is not due to rotation in our models. Instead we find that the line-of-sight velocity gradient arises primarily due to a changing orbital velocity along the trajectory of the orbit. 

\begin{figure}
  \centering \epsfxsize=8.5cm \epsffile{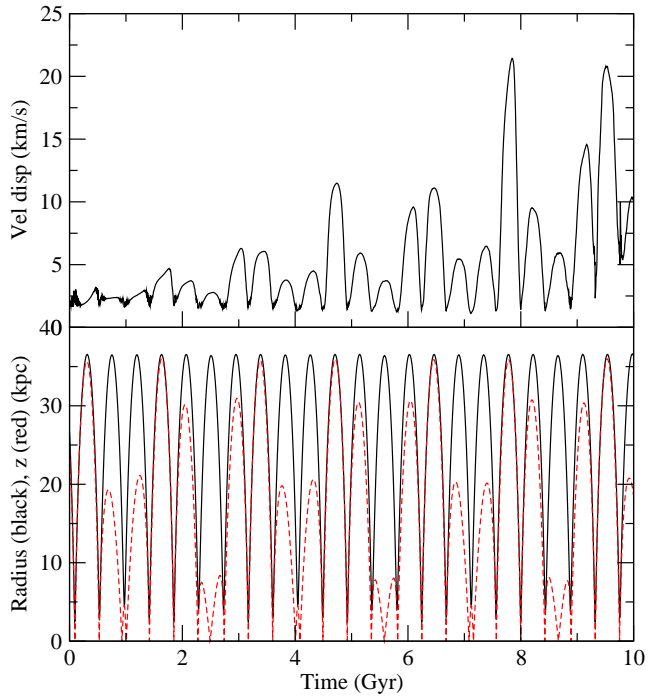}
  \caption{Fiducial Model's time evolution of; (upper panel) velocity dispersion of all stars down line-of-sight within a 1.5 by 1.5 degree square area of the remnant, (lower panel) the orbital radius (solid, black line) and z-position (dashed, red line).}
\label{boost1}
\end{figure}

\begin{figure}
  \centering \epsfxsize=8.5cm \epsffile{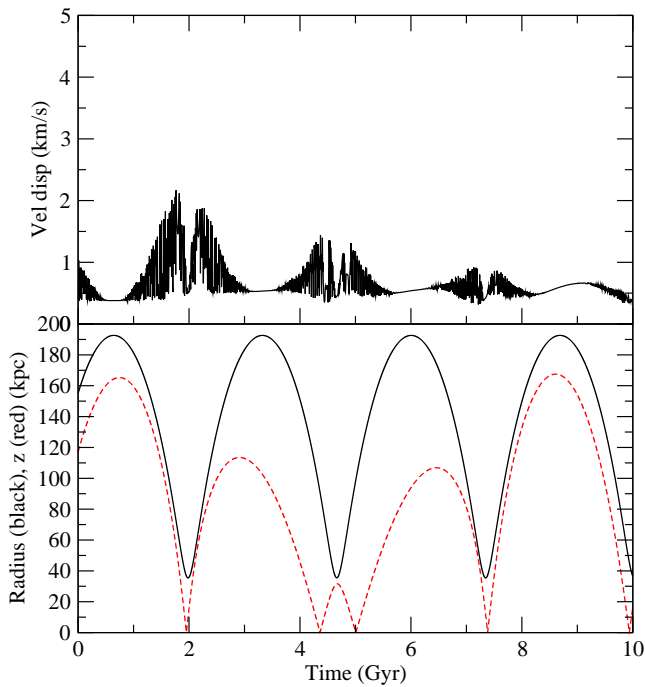}
  \caption{Same as in Fig. \ref{boost1}, but for Comparison Model 1.}
\label{boost2}
\end{figure}

\begin{figure}
  \centering \epsfxsize=8.5cm \epsffile{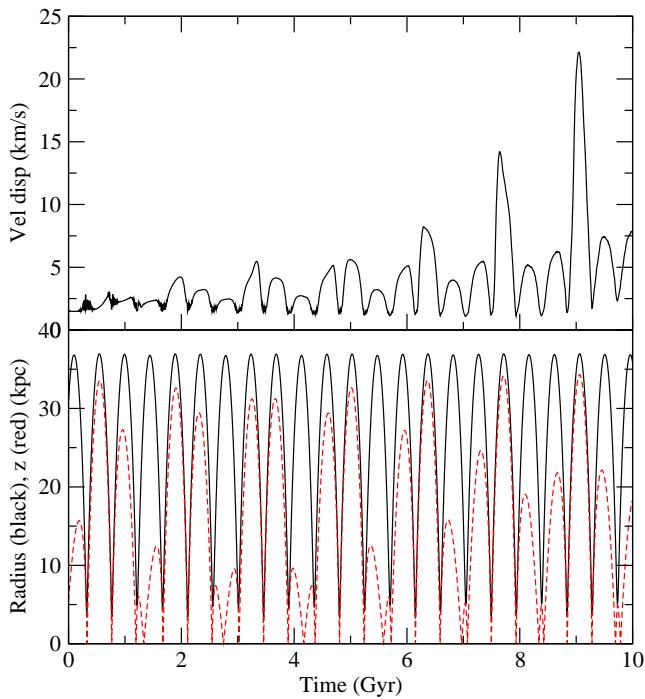}
  \caption{Same as in Fig. \ref{boost1} but for Comparison Model 2.}
\label{boost3}
\end{figure}

\subsubsection{Velocity dispersion}
\label{veldispboosts}

The upper panel of Fig. \ref{boost1} shows the time evolution of the velocity dispersion of the Fiducial Model, as might be observed by an observer in the Galacto-centric frame of reference (i.e. positioned at the centre of the MW). To calculate this, we measure the dynamics of the model particles along a 0.4~kpc \footnote{corresponding to $\sim$3 times the effective radius of UMaII} radius tube from the centre of the MW, to the model galaxy's position along its orbit at that instant, and calculate the standard deviation. The Fiducial Model clearly presents regular boosts of velocity dispersion to $>$5~km~s$^{-1}$. By comparison between the upper and lower panel, it can be seen that, for t$>$2~Gyr, these occur at each apocentre. The rounded shape of the peaks demonstrates that the velocity dispersion boosts last throughout the duration of the apocentre ($\sim$300~Myr). 

\begin{figure*}
  \centering \epsfxsize=14.cm \epsffile{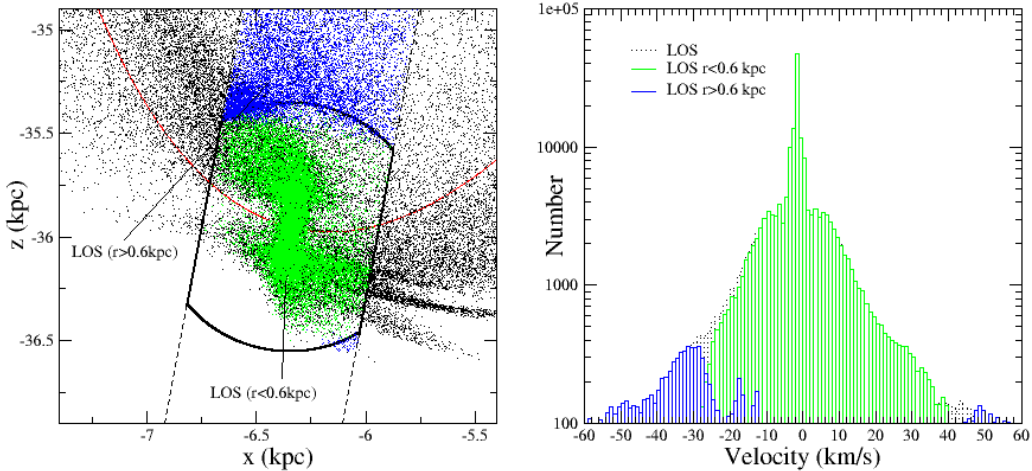}
  \caption{(left) x-z plot of particle positions of the Fiducial Model at apocentre (t=9.55~Gyr) within a 2 by 2 kpc box. Black points are all particles, particles between the dashed lines are particles down our line-of-sight within a tube of radius=0.3~kpc (`line-of-sight particles'), colour indicates if line-of-sight particles are within 0.6~kpc (green), or beyond 0.6~kpc (blue), of the centre of density of the model. The bold line is the cross-section of a sphere (radius=0.6~kpc), within the line-of-sight. (right) Histograms of the line-of-sight velocity of; all line-of-sight particles (black, dotted) all line-of-sight particles closer than 0.6~kpc (green), or further than 0.6~kpc (blue), from the centre of density.}
\label{xzboost}
\end{figure*}

It can also be seen that the size of the velocity dispersion boost is related to how close the z-distance is to the apocenter distance, i.e. whenever the z-distance approaches the apocentre distance of 36.5~kpc, a much larger boost of the velocity dispersion occurs. Orbital precession causes the model's orbital trajectory periodically to be almost perpendicular to the plane of the MW at pericentre. When this occurs, and the model reaches apocentre, the z-distance approaches the apocentre distance (e.g. see the orbital trajectory (red curve) shown in the upper panels of Fig. \ref{xzsnaps}, where the plane of the MW disk is in the x-y plane).

Apocentre boosting enables the Fiducial Model ($\sigma_{\rm{L}}=6.3~$km~s$^{-1}$) to match the observed velocity dispersion in UMaII $\sigma_{\rm{L}}$=(6.7$\pm$1.4)~km~s$^{-1}$. We note that the size of the velocity dispersion boost at apocentre is not related to the amount of mass lost in the previous pericentre, i.e. the upper panel of Fig. \ref{massloss} shows that mass loss at every pericentre pass is roughly equal.

We also measure the time evolution of the velocity dispersion down our line-of-sight for Model CM1 (Fig. \ref{boost2}) and Model CM2 (Fig. \ref{boost3}). In Model CM1, there is no sign of the substantial velocity dispersion boosting seen at apocentre in the Fiducial Model. Instead, a minor, stochastic boosting is seen at pericentre, that appears to reduce in magnitude with each pericentre passage. As a result Model CM1 ($\sigma_{\rm{L}}=0.7~$km~s$^{-1}$), falls very short of the observed velocity dispersion when it reaches UMaII's position in the sky. We note that the velocity dispersion of Model CM1 behaves similarly to the Fiducial Model and Model CM2 in the first 1-2~Gyr.

However, Model CM2 is very similar to the Fiducial Model in its velocity dispersion evolution. For t$>$2~Gyr, the velocity dispersion is boosted at each apocentre. Once more, an especially strong boost is seen whenever the z-distance approaches the apocentre distance. The size of velocity dispersion boosting is comparable to that seen in the Fiducial Model. However, purely due to timing, Model CM2 is not undergoing a large apocentre boost ($\sigma_{\rm{L}}=2.6~$km~s$^{-1}$) when it reaches UMaII's position in the sky and thus it, too, fails to match the observed velocity dispersion of UMaII.

By comparison between the models we see that the mechanism that boosts the velocity dispersion of the Fiducial Model and Model CM2, but is not seen in Model CM1, is clearly sensitive to orbit (see Tab. \ref{orbitpars}). The Fiducial Model and Model CM2 both have higher eccentricity ($\epsilon>0.8$), which may aid the contraction of the tidal streams at apocentre (as discussed in Section \ref{Fiducialevol}), perhaps enhancing velocity dispersion boosting. Model CM1 also has a much larger pericentre and apocentre, and therefore experiences weaker tides throughout its orbit. Additionally the period of Model CM1 is $\sim$6 times longer than the other models. This is likely a key cause for their differing dynamical behaviour. While Model CM1 displays many of the same features as the other models (e.g. contraction of tidal streams at apocentre, inner tidal tails), the long orbital period results in the evolution of these features occuring at a significantly more leisurely pace. A full study of the velocity dispersion boosting at apocentre, and its dependencies on orbital parameters, is beyond the scope of this study but we shall investigate this in a follow-up paper.

There is also some evidence for a gradual increase, with time, in the amplitude of the velocity dispersion peaks in the Fiducial Model and Model CM2. This is likely due to the decreasing quantity of bound mass in the model at later times, and the increasing quantities of unbound mass in the form of tidal streams at later times.

\begin{table*}
\centering
\caption{Summary of results and comparison between Ursa Major II properties with the model properties. The table is split loosely into two categories of (top) morphological parameters, and (lower) dynamical parameters.}
\begin{tabular}{|c|c|c|c|c|}
\hline
  Property & UMaII & Fiducial & Model CM1 & Model CM2\\
\hline
\hline
  V-band abs. mag. M$_{\rm{V}}$ & -3.9 $\pm$ 0.5 & -4.4 $\surd$& -3.5 $\surd$& -4.0 $\surd$\\
  Central surf. bright. $\mu_0^{\rm{V}}$ (mag~arcsec$^{-2}$)& 29.2 & 29.2$\surd$& 29.1 $\surd$& 29.1 $\surd$\\
  Ellipticity $\epsilon$ & 0.5 $\pm$ 0.2 & 0.4-0.6 $\surd$ & 0.3-0.5 $\surd$& 0.2-0.3$ \surd$\\
\hline
  Radial Velocity V$_{\rm{R}}\odot$ (km~s$^{-1}$) & -116.5 $\pm$ 1.9 & -116.4 $\surd$&-117.5 $\surd$&-118.4 $\surd$\\
  Velocity grad. (km~s$^{-1}$) & 8.4 $\pm$ 1.4 km~s$^{-1}$ & 8.4 $\surd$ & 2.8 {\bf{X}}& -4.0 {\bf{X}}\\
  Velocity disp. $\sigma_{\rm{L}}$ (km~s$^{-1}$) & (6.7 $\pm$ 1.4) & 6.3 $\surd$& 0.7 {\bf{X}}& 2.6 {\bf{X}}\\
\hline
\end{tabular}
\label{tabcomp}
\end{table*} 

To try to better understand the source of the velocity dispersion boosting at apocentre, we study a single instant of the Fiducial Model in detail. We choose t=9.55~Gyr as, at this instant, the Fiducial Model is found very close to apocentre, and presents a very strong ($\sigma=$20.7~km~s$^{-1}$) boost of velocity dispersion. In the left panel of Fig. \ref{xzboost} we show the particle distribution of the Fiducial Model, at this instant, in a 2 by 2 kpc box centred on the model. Dashed lines indicate the boundaries of the 4~kpc radius tube along which our line-of-sight velocity measurements are made. We refer to these particles as `the line-of-sight particles'. We also select a sub-set of the line-of-sight particles, that are found within 0.6~kpc of the model's centre of density (green particles). We note that this subset of particles is heavily affected by the dynamics of the inner tidal tails that point towards the MW centre, and thus along our line-of-sight (i.e. along the dashed lines). Thus we refer to these particles as `inner tidal tail particles' (although we note that particles belonging to the bound remnant are also present within this volume). Line-of-sight particles at $>$0.6~kpc from the centre of density (blue particles) tend to align themselves along the orbital trajectory of the galaxy, once they have left the inner tidal tails. We refer to these particles as `outer stream particles'. 

In the right-hand panel we plot histograms of the distribution of line-of-sight velocities for each set of particles. The line-of-sight particles are shown by black dotted lines. The inner tidal tail particles subset is shown in green, and the outer stream particle subset is shown in blue. The line-of-sight particles show a large, narrow central peak, where large numbers of particles are moving at relatively low velocities. This peak is surrounded by a wider distribution of particles. The combined narrow and wide distribution forms a double profile distribution. The narrow peak, and much of the wide distribution is dominated by the green bars of the histogram, which represent bound core stars and inner tidal tail stars. We will show in the following section that the narrow peak does not consist of bound stars alone. The inner tidal tails are aligned down our line-of-sight. Hence stars that travel towards and away from us along the tails widen the narrow distribution, and contribute to the wide distribution. The net result is that the velocity dispersion is boosted. The outer stream particles (blue histograms) only dominate the dynamics at velocities $\sim$20-30~km~s$^{-1}$ from the central peak. However, as the double profile of particle velocities is very non-Gaussian, measurement of the standard deviation of particle velocities can be additionally boosted by the presence of outer stream particles.

The combined action of the inner tidal tails and outer streams on the measured dynamics may explain why extra strong boosts occur when the z-distance is close to apocentre distance. We note that this occurs when the orbital trajectory at apocentre is most tightly curved (e.g. see the orbital trajectory in the upper panels of Fig. \ref{xzsnaps}). This could potentially enhance the boosting effect of both the inner tidal tails and outer streams, by better aligning the inner tidal tails down our line-of-sight, and causing stronger tidal stream contraction, resulting in larger numbers of outer stream stars being present in our line-of-sight. Indeed, apocentre-peaks in the number-density of stars surrounding the Fiducial Model are strongest when z-distance is close to apocentre (see middle panel of Fig. \ref{massloss}). This will be studied in detail in our follow-up paper.

A signature of this mechanism of velocity dispersion boosting at apocentre is the two distinct and superimposed velocity distributions; a narrow peaked distribution, and a second much wider distribution. We shall investigate if this two-distribution signature is detectable given measurement uncertainties and low number statistics in Section \ref{detectable}.

Summarising, in both the Fiducial Model and Model CM2 we see regular boosting of the velocity dispersion which lasts throughout the time the models are at apocentre. After 2~Gyr, this occurs at each apocentre, and therefore it is not necessary for the galaxy to be close to destruction to present an enhanced velocity dispersion (unlike in \citealp{Fellhauer2007}). Furthermore, in both the Fiducial Model and Model CM2, an especially strong apocentre boost occurs when the galaxy trajectory is near perpendicular to the plane of the MW. This occurs periodically as the orbit precesses. For the Fiducial Model, this results in a sufficiently boosted velocity dispersion to match the observed values. The mechanism behind the velocity dispersion boosting is sensitive to orbital parameters (e.g. orbital period). The stars that are responsible for velocity dispersion boosting are found in the inner tidal tails, and outer tidal streams. The inner tidal tails tend to be pointed roughly towards and away from the centre of the MW (and hence down our line-of-sight), while the galaxy is at apocentre. Also, the outer tidal streams are more contracted when the model is at apocentre, bringing more unbound stars into our line-of-sight. In the following section, we attempt to ascertain if the boosting effects of the inner tidal tails, and outer stream stars, can be filtered from the velocity dispersion of stars belonging to the bound core.

A comparison of the successes and failures of each model in matching the key observational properties of UMaII is given in Tab. \ref{tabcomp}.

\begin{figure*}
  \centering \epsfxsize=18.0cm \epsfysize=12.5cm \epsffile{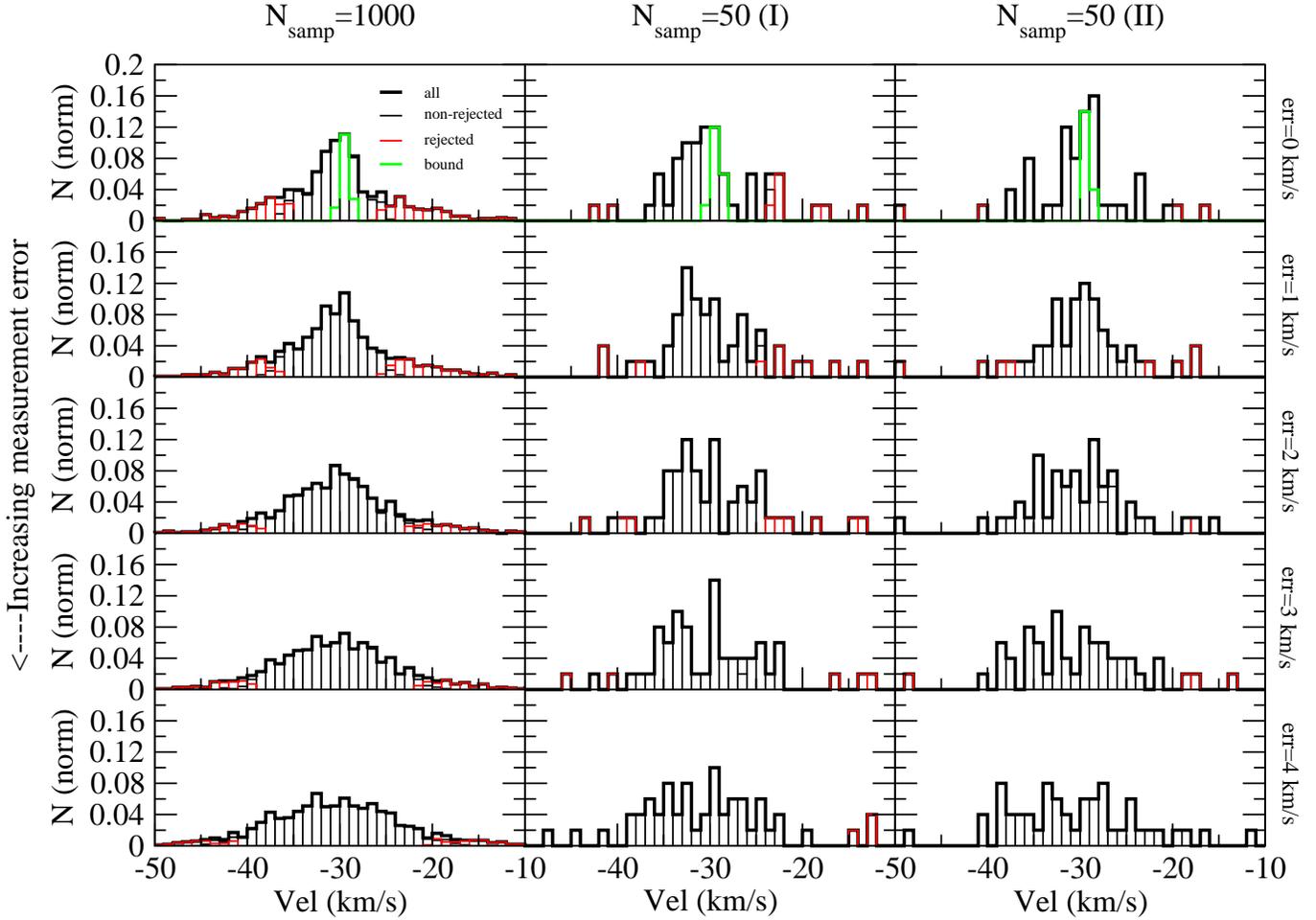}
  \caption{Normalised histograms of the line-of-sight velocities of particles in the Fiducial Model, after 10~Gyr of orbits about the MW, in a 1.5 by 1.5 degree square region centred on the remnant. Column 1 is for a single 1000 particle sample. Columns 2 and 3 are two 50 particle samples of the total number of particles, to understand the effects of low-N statistics. The top row is simply the particle velocities from the simulation. In the subsequent rows we add an error to the particle velocities, to mimic observational uncertainty in measuring stellar velocities. Rows 1, 2, 3, 4, and 5 have errors of 0, 1, 2, 3, and 4 km~s$^{-1}$.}
\label{histogramsfinal}
\end{figure*}

\section{Velocity distributions of the Fiducial Model - effects of velocity measurement uncertainties, and low number statistics}
\label{detectable}
\subsection{Detecting the double profile}

In the following we consider the velocity distribution of samples of stars in a 1.5 by 1.5 degree square (approximately 800 by 800 pc) about the remnant of the Fiducial Model after 10~Gyr of tidal disruption by the MW potential. Within this region, there are $\sim$120000 particles in total, and their velocity dispersion is 10.4~km~s$^{-1}$.

However, by using the standard deviation to calculate the dispersion, we risk mixing the velocity dispersion of bound stars, and the velocity dispersion of unbound stars together, and the combined distribution may not be well described by a Gaussian distribution.

This can be seen in the upper-left panel of Fig. \ref{histogramsfinal}. All the panels are histograms of the line-of-sight velocities of the stars in a 1.5 by 1.5 degree square about the final remnant of the Fiducial Model. The upper left panel is the result from the simulation for a randomly chosen sample of N$_{\rm{samp}}$=1000 stars from this region. We first consider the bold, black histogram bars labelled `all'. With this number of particles, the distribution can be seen to have two distinct components; a narrow distribution with low velocities ($\sim$4~km~s$^{-1}$ wide), and a second, distinct, and much wider distribution of stars ($\sim$40~km~s$^{-1}$ wide).

The green histogram bars are the line-of-sight velocities of bound stars only. The narrow distribution clearly does not consist of bound stars only. In fact, inner stream stars that are directed down our line-of-sight are also found in the central peak, causing the width of the narrow peak to be wider than that which would be produced from the bound stars alone. We will show that this challenges our attempts to separate bound stars from unbound stars in the following section. Although we cannot easily distinguish bound stars from inner stream stars in the histogram, instead it may be possible to see the double profile (narrow and wide distribution) that is shown in our models.

The overall distribution is poorly described by a single Gaussian distribution, and the standard deviation of all the stars is clearly boosted by the presence of the wide distribution of stars.  With such a large number of particles, and with effectively zero errors on our measurement of velocities, it is possible to distinguish the narrow distribution by eye from the wider surrounding distribution. However real observations of the dynamics of UMaII's stars are restricted by numbers of stars in the sample, and by uncertainty in measurement of line-of-sight velocities. The other panels in Fig. \ref{histogramsfinal} explore how this may hinder our ability to detect the double profile by eye.

For each row in Fig. \ref{histogramsfinal}, we choose a different observational uncertainty in the line-of-sight velocities. We add an offset to the measured velocity to represent the observational uncertainty. The size of the offset is drawn from a Gaussian distribution, where its standard deviation is equal to the measurement error that we choose. For the upper row we choose zero errors. For the second, third, fourth and fifth row, the error is chosen to be 1, 2, 3, and 4~km~s$^{-1}$.

In each column in Fig. \ref{histogramsfinal}, we choose a different sub-sample of the stars to investigate the effect of low-N statistics. Column 1 shows histograms for a sample of 1000 of all the particles in a 1-degree square region surrounding the remnant. Hence in this column, effects of low-N statistics are very reduced. However in column 2 we randomly select a sample of 50 particles from the total number to create a small sample. We select another 50 stars randomly to make a second small sample (shown in column 3). 

First we consider the panels in column 1. Once more we consider the bold, black histogram bars labelled `all'. With zero observational errors, and 1000 particles, the distribution of bound particles can be distingushed from the other distribution by eye. However, even with only 1~km~s$^{-1}$ errors, separating the two distributions becomes very difficult. Thus even with large samples of particles, errors in velocity measurements must be extremely small to enable easy separation of the distributions of bound and unbound stars. Now we consider the effects of low-N in the upper row. With only 50 particles, separating the two distributions is further compounded, and is very difficult even with small measurement errors.

\subsection{Detecting the bound stars using an interloper rejection technique (IRT)}
We test the success of the interloper rejection technique (IRT) described in \cite{Klimentowski2007}, in filtering out unbound stars for our model of UMaII. This technique has been shown to be successful in removing 70-80$\%$ of the unbound stars in a dSph model with an NFW dark matter halo, and exponential disk of stars, while only rejecting 1$\%$ of the bound stars (\citealp{Klimentowski2007}). This technique has been applied by \cite{Lokas2009} to large kinematic samples for the Fornax, Carina, Sculptor and Sextans dwarfs (data published by \citealp{Walker2009}).

To find interloper particles, first an estimated mass profile is calculated using the standard mass estimator $M_{\rm{VT}}$, derived from the virial theoreom (Equation 7 of \citealp{Klimentowski2007}). The method then assumes that a tracer particle can either be on a circular orbit with velocity $v_{\rm{circ}}$=$\sqrt{G M(r)/r}$ or falling freely in the galaxy's potential with velocity $v_{\rm{inf}}$=$\sqrt{2} v_{\rm{circ}}$ (see \citealp{denHartog1996}). Finding the maximum of the two for a given projected radius $R$, all stars with greater velocities are rejected from the sample. Now a new estimated mass profile is constructed with the new sample, and the process is repeated iteratively. For a more detailed description please see \cite{Klimentowski2007} and references therein.

We apply the technique to the line-of-sight velocities of our fiducial model, when it reaches UMaII's current position in the sky. The colours of the bars in Fig. \ref{histogramsfinal} indicate whether they are rejected stars (red bars), or non-rejected stars (thin black bars). We also show the velocities of the bound stars as green bars.

With zero measurement errors and N$_{\rm{samp}}$=1000 (upper-left panel), the IRT effectively rejects the majority of the stars in the wide distribution. In the narrow peaked distribution ($\sim$4~km~s$^{-1}$ wide), we find a combination of bound stars (green) and unbound stars. This illustrates that the central peak is not purely a result of the low velocities of bound stars, but also has a considerable contribution from unbound stars within the inner tidal tails. The IRT fails to reject these unbound stars (e.g. compare the width of the non-rejected stars and bound stars). 

This failure is due to an overestimate of the bound mass, calculated using the standard mass estimator $M_{\rm{VT}}$. This occurs due to the dynamics on the inner tidal tails down our line-of-sight, and the mass is heavily overestimated (by factor of $>$1000). The mass is overestimated by a similar factor, even if we use all the simulation particles available in our field of view ($\sim$1$\times$10$^5$ particles). Besides overestimating the mass, the IRT fails to reject unbound stars. For example, the IRT finds 64.1$\%$ as non-rejected for zero errors, when actually only 15.6$\%$ are bound within our field-of-view -- over counting the number of bound particles by a factor of f$_{\rm{nonreject}}$/$f_{\rm{bound}}$=4.1. The degree of failure increases with increasing measurement error (moving to lower rows in Fig. \ref{histogramsfinal}), with the failure increasing roughly linearly to f$_{\rm{nonreject}}$/$f_{\rm{bound}}$=5.1 for measurement errors of 4~km~s$^{-1}$ (corresponding to a mass overestimate of $>$7000). 

We repeat this analysis for the two 50 particle samples (second and third column of Fig. \ref{histogramsfinal}). We find similar behaviour with f$_{\rm{nonreject}}$/$f_{\rm{bound}}$=4.0-4.6 for the first sample, and f$_{\rm{nonreject}}$/$f_{\rm{bound}}$=4.2-4.9 for the second sample. Therefore even with a large sample of 1000 stars, and with zero measurement error, we find it is difficult to filter out the unbound stars, in order to remove the dynamical effects of the inner tidal tails.

In summary, we find that the velocity distribution of our remnant galaxies consist of a narrow distribution of bound and inner tidal tail stars, surrounded by a wide distribution of inner tidal tail and outer stream stars. We refer to the combined narrow and wide distribution as a double profile. Tidal stripped stars are responsible for boosting the standard deviation measurements when the remnants are near apocentre. We find that with large particle samples (N$_{\rm{samp}}$=1000), and no errors on velocity measurements, we can distinguish the narrow and wide distributions of stars by eye. This is even easier when a more massive bound core of particles remains (i.e. at early times when the remnant is not close to disruption). However the uncertainties in measuring the velocities of real stars can quickly remove any trace of the double profile -- and this occurs with only 1-2~km~s$^{-1}$ errors. Therefore if tidal effects are indeed responsible for the observed high velocity dispersion of UMaII, we conclude the following: {\it{only very high precision measurements ($\sim$1~km~s$^{-1}$ errors) of stellar velocities could enable us to detect the double profile signature resulting from the unbound stars.}} However, even if the double profile is detected, the narrow distribution consists of both bound core stars and unbound, inner tidal tail stars. Therefore the narrow peak's velocity distribution does not accurately reflect the dynamics of the bound stars.

\subsection{Evolution of mass and velocity dispersion methods with time - the interloper rejection method versus sigma-clipping}

\begin{figure*}
  \centering \epsfxsize=17.5cm \epsfysize=19.0cm \epsffile{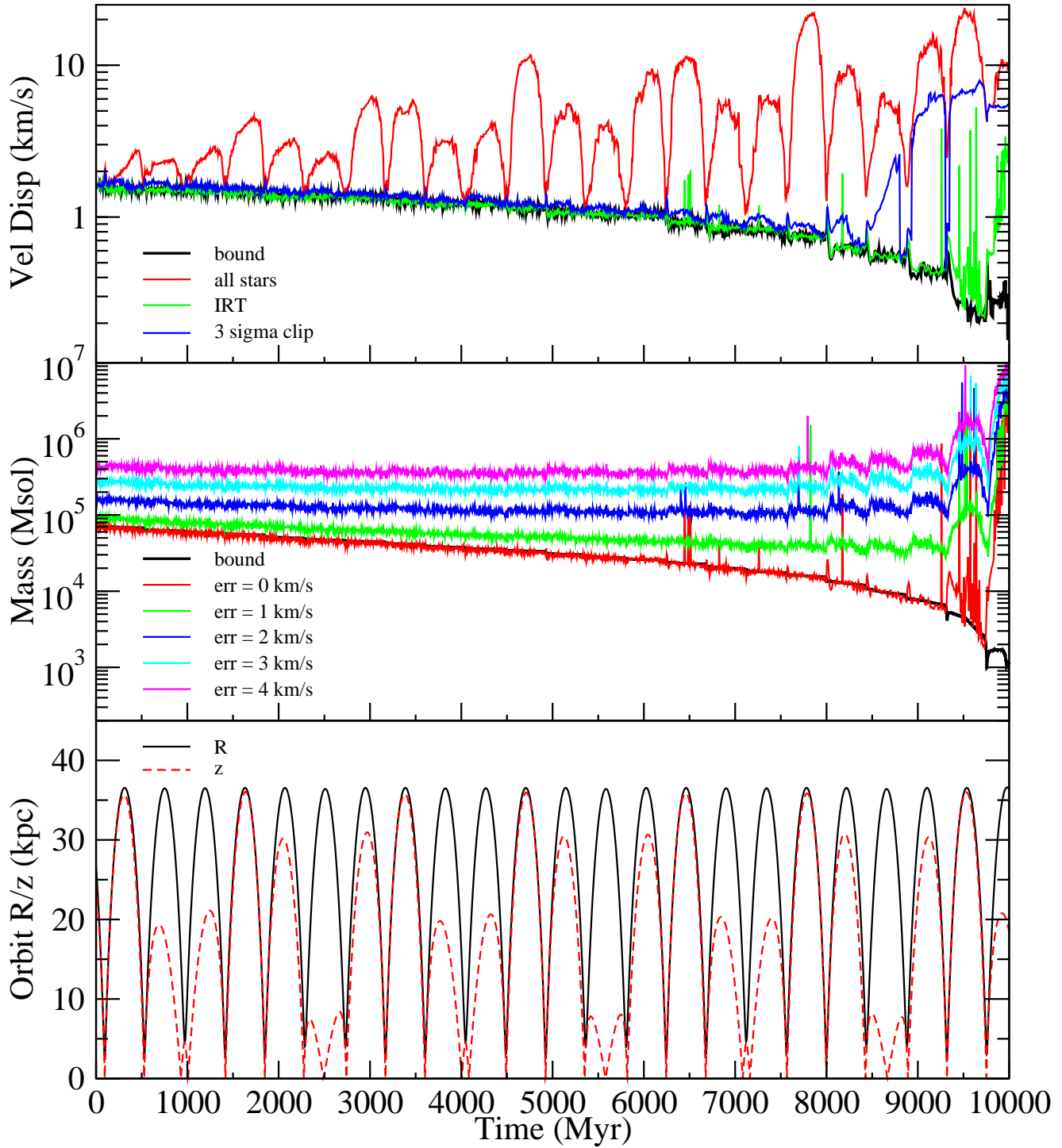}
  \caption{(Upper panel) Time evolution of the velocity dispersion of the Fiducial Model for; bound stars (black line), all stars of 1000 particle sample (red line), remaining stars after using the interloper rejection technique (green line), and remaining stars after using an interative 3-sigma clip (blue line). The interloper rejection technique (IRT) and 3-sigma clip effectively removes the effects of apocentre velocity dispersion boosting until late stages. Then the 3-sigma clip fails first, at t$\sim$8.5~Gyr). Meanwhile the IRT only fails substantially at later times (t$\sim$9.5~Gyr). (Middle panel) Time evolution of the remaining mass of the Fiducial Model as measured by counting the number of bound stars (black, bold line), or using the interloper rejection technique with varying quantities of measurement error (line colours presented in key show size of measurement error). With zero measurement error, the IRT effectively measures the remaining bound mass until the final 500~Myr. However with measurement errors, the interloper rejection technique always overestimates the remaining mass, and only detects the reduction in the mass for errors $\le$1~km~s$^{-1}$. (Lower panel) The orbital radius $R$ (black, solid line) and the absolute z-distance from the plane of the disk $z$ (red, dashed line)}
\label{filtertest}
\end{figure*}

We note that the final remnant of the Fiducial Model is perhaps a worst-case scenario for detecting the double-profile, and for using the IRT, as by t=10~Gyr the mass of the bound core is very low (e.g. see lower panel of Fig. \ref{massloss}). It is therefore important to consider the Fiducial Model at earlier times, when a larger mass bound core exists.

We test the success of the IRT, at measuring the velocity dispersion and mass (upper and middle panel of Fig. \ref{filtertest} respectively) of the bound core of the Fiducial Model, throughout the 10~Gyr duration of the simulation. In practice we randomly choose a 1000 particle sample of stars from a 1.5 degree square area surrounding the Fiducial model, and measure the velocity dispersion and mass from these stars using the interloper rejection technique. For comparison, the lower panel shows the orbital radius $R$, and absolute z-distance from the plane of the disk $z$ at any instant. 

The velocity dispersion of the bound stars is shown by the black, bold curve in the upper panel of Fig. \ref{filtertest}. The velocity dispersion of all the stars in the 1000 particle sample is shown by the red curve. The velocity dispersion boosting at apocentre is visible, and is similar to that shown in Fig. \ref{boost1}. The velocity dispersion measured using the IRT is shown by the green curve. The IRT has effectively filtered out all stars that result in the apocentre boosting. Only a very minor boosting ($\sim$0.2~km~s$^{-1}$) occurs periodically at pericentre after 7.5~Gyr. However, in the final 0.5~Gyr, when only $\sim$2$\%$ of the initial mass remains bound, the IRT starts to fail, returning velocity dispersions that are boosted. We also compare the success of the interloper rejection technique to a simple iterative three-sigma clip (\citealp{Yahil1977}) as shown by the blue curve. Similarly, at early times the sigma-clipping filters out the velocity dispersion boosting at apocentre seen in the `all stars' sample. However the velocity dispersion boosting at apocentre returns at t$>$8.5~Gyr, and results in heavily enhanced velocity dispersion measurements ($\sim$7-8~km~s$^{-1}$), when 15$\%$ of the UMaII model remains bound. Therefore the IRT provides more reliable velocity dispersion measurements than sigma-clipping, and for longer -- only failing in the final 0.5~Gyr.

In the middle panel, the bound mass of the Fiducial Model is shown by the black bold line. Each coloured line is the measured mass using the IRT, but with differing measurement error on line-of-sight velocities (0-4~km~s$^{-1}$ -- see key). As previously, we add an offset to the measured velocity to represent the observational uncertainty. Once more, the size of the offset is drawn from a Gaussian distribution, where its standard deviation is equal to the measurement error that we choose. In the case of zero measurement error, we see that the mass measured using the IRT is extremely reliable until the final 0.75~Gyr of the simulation, when $<$9$\%$ of the model remains bound. Then the measured mass can be boosted by greater than two orders of magnitude, due to the dynamics of the inner tidal tails. With non-zero measurement error we see that, right from the beginning, all measured masses are raised by a roughly constant factor (e.g. $\sim$1.3 for 1~km~s$^{-1}$ errors, $\sim5.7$ for 4~km~s$^{-1}$ errors). This can be understood by considering the following extreme case. Even if all the bound stars were to have an identical line-of-sight velocity (i.e. with 0~km~s$^{-1}$ errors), then with 4~km~s$^{-1}$ errors they will appear to have a gaussian velocity distribution with a standard deviation of 4~km~s$^{-1}$. The measurement errors thus widen the velocity distribution of all the stars -- even the bound stars. This has a significant effect, resulting in the IRT overestimating the measured masses. This mass overestimate can be severe. For measurement errors $>$1~km~s$^{-1}$, the IRT fails to detect {\it{any}} of the mass loss suffered by the Fiducial Model over the 10~Gyr duration of the simulation. Even for very modest measurement errors of only 1~km~s$^{-1}$, at t=9~Gyr, the IRT finds the mass has reduced by only $\sim$50$\%$, when it finds 90$\%$ mass reduction for zero measurement error.

In summary, for t$<$8.5~Gyr, we find that 3-sigma clipping method and the interloper rejection technique (IRT) can effectively remove the velocity dispersion boosting previously seen at each orbital apocentre. However when $<$15$\%$ of the model remains bound, the sigma-clipping provides highly boosted velocity dispersions at apocentre. The IRT is an improvement, and provides reliable velocity dispersions until $<$2$\%$ of the model remains bound. In the case of zero measurement errors on line-of-sight velocities, the IRT also provides very reliable mass measurements of the bound core until $<$9$\%$ of the model remains bound, and only becomes highly unreliable when $<$2$\%$ of the model remains bound. Therefore, in general the IRT is found to be a highly successful approach to measuring the bound core's mass and velocity dispersion. However, when the galaxy is nearing destruction ($<$10$\%$ of the initial mass remains bound), the dynamics of the inner tidal tails become relevant, and it is difficult to remove their influence from mass and velocity dispersion measurements of the bound core. Even with zero measurement error, and more than one hundred thousand velocity measurements, it is not possible to distinguish the dynamics of bound stars from inner tidal tail stars using the IRT. As a result the IRT does not reject unbound, inner tidal tail stars, and heavily overestimates the bound mass by several orders of magnitude. Finally, we find that even with very modest measurement errors on line-of-sight velocities ($\sim$1~km~s$^{-1}$), the measurement of the mass of the model is raised by an approximately constant factor from the true mass. In practice we find that this causes masses derived using the IRT to be overestimated. This is quite substantial, and for measurement errors of $>$1~km~s$^{-1}$, we find that the IRT completely fails to detect the decreasing bound mass of the Fiducial Model over the duration of the simulation -- even when it loses as much as $\sim97\%$ of its bound mass.

\section{Summary $\&$ Conclusions}
Using N-body simulations, we evolve three progenitor models along three orbits in the potential well of the MW. Over 10 Gyrs of each orbit, the models are heavily tidally stripped, under the influence of MW gravitational tides. We test if we can reproduce the observed properties of UMaII, as a result of tidal mass loss, for three different progenitor models. 

For a direct quantitive comparison between UMaII observations, and the results of our models, please see Tab. \ref{tabcomp}. We find we can well reproduce the majority of the morphological properties of UMaII including: luminosity, central surface brightness, ellipticity, and distorted appearance for all three models (despite very different orbits and progenitor properties). However, to better match the surface brightness profile at small radii, we would require a progenitor model with a more cuspy profile.

We also find that the observed radial velocity gradient can also be reproduced by the MW potential, although in this scenario it is not an indicator of rotation. In fact it is a natural result of the acceleration (or de-acceleration) of unbound stars that slightly lead or trail the model as it falls into (or climbs out of) the potential well of the MW. Therefore producing the velocity gradient in this manner requires greater constraints on orbit than are required to reproduce the UMaII morphology. As such, of our three models, only the Fiducial Model orbit produces sufficent orbital acceleration, at UMaII's current position in the sky, to reproduce the observed velocity gradient. 

Additionally we find that our models' line-of-sight velocity dispersion can be raised to the observed values in UMaII. This form of velocity dispersion boosting occurs for an extended duration, when the model is close to apocentre. For $t>$2~Gyr, at each apocentre the observed velocity dispersion is raised to $>$5~km~s$^{-1}$. Thus the galaxy need not be close to complete destruction in order to display an enhanced velocity dispersion, unlike in \cite{Fellhauer2007}. We also see even stronger boosting of the velocity dispersion that occurs periodically. This occurs when orbital precession causes the model trajectory to be close to perpendicular to the plane of the MW disk at pericentre, and can result in velocity dispersions of $>$20~km~s$^{-1}$ when the model arrives at apocentre. This form of velocity dispersion boosting is primarily responsible for the Fiducial Model being able to match the high observed velocity dispersions in UMaII. These results demonstrate that tidal effects are sufficient to mimic the influence of a dominating, massive dark matter halo on stellar velocity dispersions -- even when there is no dark matter present in any of our models. We do not believe that these simulations cast any light on the past, or current, dark matter content of UMaII. Rather they highlight the considerable uncertainty that exists in current measurements of the dark matter content of UMaII.

A velocity distribution histogram of our models reveals a signature two profile distribution of stars; a narrow peaked distribution consisting of both bound and inner tidal tail stars, surrounded by a wider distribution of inner tidal tail stars, and outer stream stars. The velocity distribution is therefore poorly described by a Gaussian distribution. It is the presence of the inner tidal tail, and outer stream stars that causes the velocity dispersion boosting. The inner tidal tails are aligned down our line-of-sight when the model is at apocentre. Also, the outer stream stars are more contracted at apocentre, bringing more unbound stars into our line-of-sight. Therefore it is the dynamics of tidally unbound stars that dominate the velocity dispersion at apocentre. However with uncertainties in the measurement of the velocities of stars of only 1-2~km~s$^{-1}$, we find it is very challenging to distinguish the signature double profile, by eye, in velocity histograms. This suggests that very high accuracy velocity measurements ($\sim$1~km~s$^{-1}$) are required to distinguish between the stellar dynamics produced in our scenario, and that of bound stars in a significantly more massive dark matter halo. 

We also test the success of the iterative interloper rejection technique (IRT) of \cite{Klimentowski2007} in identifying unbound stars in our Fiducial Model, and compare it to a more traditional three-sigma clip (\citealp{Yahil1977}). We find the IRT is very successful at removing the velocity dispersion boosts we see at apocentre, and returns reliable mass measurements until only $\sim$10$\%$ of the initial mass remains bound. However when $<$2$\%$ of the mass remains bound, the dynamics of the inner tidal tails causes the IRT to massively over-estimate the bound mass (by factors of $>$1000). This is the case when our model reaches UMaII's current location in the sky. The narrow distribution of the double profile consists of bound stars and unbound, inner tidal tail stars. However, the IRT fails to remove the unbound stars from the narrow distribution, and as a result they enhance the velocity width of the narrow distribution, causing the substantial mass overestimate. Measurements of the dark matter content of UMaII are calculated based on the assumption of dynamical equilibrium between the tracer population and the galaxy's potential. The failure of the IRT to remove these unbound stars underlines how incorrect the assumption of dynamical equilibrium may be. 

We also find that mass measurements are sensitive to the measurement errors on the line-of-sight velocities of stars. For measurement errors of 2-4~km~s$^{-1}$, masses are typically overestimated by a factor of 2-6, purely due to the measurement error. The IRT mass overestimate can be severe. For example, with only 2~km~s$^{-1}$ measurement uncertainties, the IRT measures the mass at any instant to be roughly constant, even when in reality $\sim97\%$ of the mass is unbound by the end of the simulation.

Our key results may be summarised as follows. 

\begin{enumerate}
\item Tidal mass loss enables us to reproduce the observed luminosity, central surface brightness, distorted morphology and eccentricity of UMaII and, with the correct choice of progenitor, this can be accomplished for widely differing orbits.
\item The potential well of the galaxy also enables us to reproduce the observed velocity gradient across UMaII. In our scenario, the gradient is unrelated to rotation, and occurs due to the orbital acceleration of unbound stars ahead and behind of the model, as they climb out (or fall into) the potential well of the MW.
\item The dynamics of unbound stars can boost the observed velocity dispersion to $>$5~km~s$^{-1}$. This occurs for extended periods each time the model is near apocentre. It is not a requirement that the galaxy be close to disruption to produce a boosted velocity dispersion. The strength of the boosting at apocentre appears to be related to the angle of inclination of the orbital trajectory relative to the MW disk. When the orbital trajectory is $\sim$90 degrees to the MW disk (at pericentre), the boosting (at apocentre) is strongest (e.g. $>$20~km~s$^{-1}$).
\item The velocity distribution of our models consists of a double profile; a narrow peaked distribution, surrounded by a much wider distribution of stars. The narrow distribution consists of bound core stars and inner tidal tail stars, while the wide distribution consists of inner tidal tail stars and outer stream stars. It is the inner tidal tail stars, and outer stream stars that are responsible for the velocity dispersion boosting at apocentre. We find that it is almost impossible to distinguish these two distributions by eye in velocity distribution histograms with $>$2~km~s$^{-1}$ errors in velocity measurements. Therefore high precision velocity measurements are required to test our scenario.
\item We find the iterative interloper rejection technique (IRT), described in \cite{Klimentowski2007}, is a substantial improvement over the more standard technique of three-sigma clipping, providing very reliable measurments of the velocity dispersion and mass of our Fiducial Models's bound stars while $>$10$\%$ of the initial mass is still bound.
\item However, when the Fiducial Model has $<$2$\%$ of its initial mass still bound, the technique tends to heavily overestimate the bound mass by a factor $>$1000. This is due to the dynamics of the inner tidal tails, and this occurs when the model reaches UMaII's current location in the sky. The IRT fails to reject unbound, inner tidal tail stars from the narrow distribution of the double profile, and the presence of these unbound stars causes the significant enhancement of the mass measurement.
\item We also find that IRT mass measurements are sensitive to measurement error on line-of-sight velocities, increasing measured masses by factors of 1.3-5.7 for errors of 1-4~km~s$^{-1}$. For $\ge$2~km~s$^{-1}$ errors, we find the IRT is incapable of detecting the mass reduction from tidal mass loss, despite the loss of $>$90$\%$ of the bound mass of the Fiducial Model.
\end{enumerate}

\section*{Acknowledgements}
MF acknowledges support by FONDECYT grant 1095092 and FONDECYT grant 1130521 , RS acknowledges support by FONDECYT grant 3120135, and GC acknowledges support by FONDECYT grant 3130480. Dark Cosmology Centre is funded by the DNRF. RS thanks Jaroslaw Klimentowski for his help. The authors acknowledge the referee for helpful feedback and suggestions that have overall improved the paper.
\bibliography{bibfile}

\bsp

\label{lastpage}

\end{document}